%% file: main.tex
\documentclass[onecolumn,twoside]{IEEEtran}

\usepackage{times}
\usepackage{amsmath}  
\usepackage{amssymb}  
\usepackage{theorem}  
\usepackage{cite}     
\usepackage{upref}
\usepackage{amsfonts}
\usepackage{verbatim}
\usepackage[dvipsnames,usenames]{color}
\usepackage{cite}
\usepackage[inline]{enumitem}
\usepackage{caption}
\usepackage{balance}
\usepackage{nicefrac}

\usepackage{algorithm}
\usepackage{times,color}
\usepackage{enumitem}
\usepackage{float}
\usepackage{color}
\usepackage{mathtools}
\usepackage{graphicx}
\usepackage{listings}
\usepackage{tikz}
\usepackage{multicol,multirow}
\usepackage{psfrag}
\usepackage{bbm}
\usepackage{verbatim}
\usepackage{soul}

\usepackage{pgfplots}
\usepackage{hhline}
\usepackage{xcolor,colortbl}
\usepackage{algorithm}
\usepackage[noend]{algpseudocode}

\usepackage{enumitem}

\usetikzlibrary{%
	arrows,%
	automata,%
	calc,%
	patterns,%
	plotmarks,%
	positioning,%
	shapes.geometric,%
	shapes,%
	snakes,%
	decorations.pathmorphing,%
	decorations.shapes,%
	graphs,%
	chains,%
	arrows.meta, %
}

\algnewcommand{\Initialize}[1]{%
	\State \textbf{Initialize:}
	\Statex \hspace*{\algorithmicindent}\parbox[t]{.8\linewidth}{\raggedright #1}
}

\input{commands}

\begin{document}

\title{\textbf{Criss-Cross Insertion and Deletion Correcting Codes}}%
\author{%
	\IEEEauthorblockN{{\textbf{Rawad Bitar}}, 
	            {\textbf{Lorenz Welter}}, 
				{\textbf{Ilia Smagloy}}, 
                {\textbf{Antonia Wachter-Zeh}}, 
		and {\textbf{Eitan Yaakobi}}
		} \\
	\thanks{RB, LW and AW-Z are with the Institute for Communications Engineering, Technical University of Munich (TUM), Germany. Emails: \{rawad.bitar, lorenz.welter,antonia.wachter-zeh\}@tum.de.}
	\thanks{IS and EY are with the CS department of Technion --- Israel Institute of Technology, Israel. Emails: \{ilia.smagloy, yaakobi\}@cs.technion.ac.il.}
	\thanks{Preliminary results of this work \cite{bitar2020criss} will be presented at ISITA, 2020.}
	\thanks{This project has received funding from the European Research Council (ERC) under the European Union’s Horizon 2020 research and innovation programme (grant agreement No. 801434) and from the Technical University of Munich - Institute for Advanced Studies, funded by the German Excellence Initiative
and European Union Seventh Framework Programme under Grant Agreement
No. 291763.}\vspace{-4ex}
}

\maketitle

\begin{abstract}
This paper studies the problem of constructing codes correcting deletions in arrays. Under this model, it is assumed that an $n\times n$ array can experience deletions of rows and columns. These deletion errors are referred to as \emph{$(\tr,\tc)$-criss-cross deletions} if $\tr$ rows and $\tc$ columns are deleted, while a code correcting these deletion patterns is called a \emph{$(\tr,\tc)$-criss-cross deletion correction code}. The definitions for \emph{criss-cross insertions} are similar. 

It is first shown that when $t_r=t_c$ the problems of correcting criss-cross deletions and criss-cross insertions are equivalent. The focus of this paper lies on the case of $(1,1)$-criss-cross deletions. \ExtVe{A non-asymptotic upper bound on the cardinality of $(1,1)$-criss-cross deletion correction codes is shown which assures that the redundancy is at least $2n-3+2\log n$ bits.} A code construction with an existential encoding and an explicit decoding algorithm is presented. The redundancy of the construction is at most $2n+4 \log n + 7 +2 \log e$. 
\ExtVe{A construction with explicit encoder and decoder is presented. The explicit encoder adds an extra $5\log n + 5$ bits of redundancy to the construction.}
\end{abstract}

\section{Introduction} \label{sec:intro}
\input{intro}

\section{Definitions and Preliminaries}\label{sec:def}
\input{defs}

\section{Main Results}\label{sec:results}
\input{results}
\section{Equivalence between Insertion and Deletion Correction}\label{sec:equiv}
\input{equivalence.tex}

\section{Upper Bound on the cardinality}\label{sec:bounds}
\input{bounds}

\section{Construction}\label{sec:cons}
\input{cons}
\section{Construction with Explicit Encoder}\label{sec:explicit}
\input{systematic}

\section{Conclusion}\label{sec:concl} \input{conclusion.tex}

\section{Acknowledgements}
\reply{We thank the associate editor and the anonymous reviewers for their valuable comments that contributed to the improvement of the quality of this work.}

\bibliographystyle{IEEEtran}
\bibliography{refs}

\appendices
\section{Proof of Claim~\ref{claim:insertion}} \label{app:claim_ins}
\input{claim_ins}
\section{Proof of Corollary~\ref{cor:upper_bound}} \label{app:cor_asymptotic}
\input{cor_asymp}
\section{Proofs of Claim~\ref{claim:rect} and Claim~\ref{claim:square}}\label{app:cons_claims}
\input{cons_claims.tex}
\section{Proof of Claim~\ref{claim:u5}}
\label{app:claim_u5}
\input{u5_claim}
\balance
\end{document}

%% file: commands.tex
\newcommand{\reply}[1]{#1}

\newcommand{\ie}{{i.e., }}

\newcommand{\be}[1]{\begin{equation}\label{#1}}
\newcommand{\ee}{\end{equation}}

\newcommand{\bc}{\begin{center}}
\newcommand{\ec}{\end{center}}

\newcommand{\cB}{{\cal B}}
\newcommand{\cC}{{\cal C}}

\newcommand{\cG}{{\cal G}}

\newcommand{\cI}{{\cal I}}

\newcommand{\cO}{{\cal O}}
\newcommand{\cP}{{\cal P}}

\newcommand{\cS}{{\cal S}}
\newcommand{\cT}{{\cal T}}
\newcommand{\cU}{{\cal U}}
\newcommand{\cV}{{\cal V}}

\newcommand{\bfc}{{\boldsymbol c}}

\newcommand{\bfr}{{\boldsymbol r}}
\newcommand{\bfs}{{\boldsymbol s}}

\newcommand{\bfB}{{\mathbf B}}
\newcommand{\bfC}{{\mathbf C}}
\newcommand{\bfD}{{\mathbf D}}

\newcommand{\bfI}{{\mathbf I}}

\newcommand{\bfU}{{\mathbf U}}
\newcommand{\bfV}{{\mathbf V}}

\newcommand{\bfX}{{\mathbf X}}
\newcommand{\bfY}{{\mathbf Y}}
\newcommand{\bfZ}{{\mathbf Z}}
\newcommand{\rIndex}{{i_R}}
\newcommand{\cIndex}{{i_C}}
\newcommand{\iX}{{\bfI^x}}
\newcommand{\iY}{{\bfI^y}}
\newcommand{\xMidDel}{{\bfX^{-1,0}}}
\newcommand{\yMidDel}{{\bfY^{0,-1}}}
\newcommand{\xMidIns}{{\bfX^{1,0}}}
\newcommand{\yMidIns}{{\bfY^{0,1}}}

\newcommand{\xij}{{X_{i,j}}}
\newcommand{\yij}{{Y_{i,j}}}
\newcommand{\dij}{{D_{i,j}}}

\newcommand{\ixij}{{I^x_{i,j}}}
\newcommand{\iyij}{{I^y_{i,j}}}

\renewcommand{\leq}{\leqslant}

\renewcommand{\geq}{\geqslant}

\newcommand{\N}{\mathbb{N}}

\newcommand{\Cref}[1]{Co\-rol\-la\-ry\,\ref{#1}}

\newtheorem{theorem}{Theorem} 
\newtheorem{lemma}[theorem]{Lemma} 
\newtheorem{lem}[theorem]{Lemma}
\newtheorem{definition}{Definition} 
 
\newtheorem{cor}[theorem]{Corollary}

\newtheorem{claim}[theorem]{Claim}
\newtheorem{proposition}[theorem]{Proposition}

\newtheorem{example}{Example}

\newtheorem{construction}{Construction}

\definecolor{Codecolor}{named}{White}  

\newcommand{\Copen}{\mbox{\{\kern-5.50pt\{}}
\newcommand{\Cclose}{\mbox{\}\kern-5.50pt\}}}
\newcommand{\Cslash}{\mbox{$\backslash\kern-6.02pt\backslash$}}

\newcommand{\oi}{\smash{\Sigma}}

\newcommand{\sigmatn}{\smash{\Sigma ^{n \times n} }}

\newcommand{\sigmatnq}{\smash{\Sigma ^{n \times n} _q}}
\newcommand{\sigmatniq}{\smash{\Sigma ^{(n-1) \times (n-1)}_q}}

\newcommand{\dball}{\ensuremath{\mathbb{D}}}
\newcommand{\iball}{\ensuremath{\mathbb{I}}}

\newcommand{\tr}{\ensuremath{t_{\mathrm{r}}}}
\newcommand{\tc}{\ensuremath{t_{\mathrm{c}}}}

\newcommand{\XR}{\ensuremath{\bfX_{\mathrm{R}}}}
\newcommand{\XT}{\ensuremath{\bfX_{\mathrm{T}}}}
\newcommand{\XL}{\ensuremath{\bfX_{\mathrm{L}}}}
\newcommand{\XB}{\ensuremath{\bfX_{\mathrm{B}}}}
\newcommand{\XC}{\ensuremath{\bfX_{\mathrm{C}}}}

\newcommand{\Xr}[1]{\ensuremath{\bfX_{\mathrm{R}#1}}}
\newcommand{\Xt}[1]{\ensuremath{\bfX_{\mathrm{T}#1}}}
\newcommand{\Xl}[1]{\ensuremath{\bfX_{\mathrm{L}#1}}}
\newcommand{\Xb}[1]{\ensuremath{\bfX_{\mathrm{B}#1}}}
\newcommand{\Xc}[1]{\ensuremath{\bfX_{\mathrm{C}#1}}}

\newcommand{\XI}[1]{\ensuremath{\bfX_{\mathrm{I}#1}}}
\newcommand{\Xii}[1]{\ensuremath{\bfX_{\mathrm{II}#1}}}
\newcommand{\Xiii}[1]{\ensuremath{\bfX_{\mathrm{III}#1}}}
\newcommand{\Xiv}[1]{\ensuremath{\bfX_{\mathrm{IV}#1}}}

\newcommand{\imin}{\ensuremath{i_{\min}}}
\newcommand{\imax}{\ensuremath{i_{\max}}}

\def \codename{CrissCross }
\def \crisscross{$(1)$-criss-cross }
\newcommand{\VTab}[1]{\cV\cT_{n,#1}(a,b)}
\newcommand{\VTcd}[1]{\cV\cT_{n-1,#1}(c,d)}
\newcommand{\VTs}[1]{\cV\cT^{\star}_{n,#1}}
\newcommand{\VTabs}[1]{\cV\cT^{(h)}_{n,#1}}
\newcommand{\VTcds}[1]{\cV\cT^{(v)}_{n-\ell-1,#1}}
\newcommand{\bfXi}[1]{\bfX^{(#1)}}

\newcommand{\ExtVe}[1]{\textcolor{black}{#1}}

\def \th{\text{-th}}

\def \Alt{\mathbf{W}}


%% file: intro.tex
\ExtVe{Recently, codes correcting insertions/deletions attracted a lot of attention due to their relevance in many applications such as DNA-based data storage systems~\cite{Heckel_A-Characterization-of-the-DNA-Storage-Channel_19}, communication systems~\cite{DolecekAnan_Sync_2007} and file synchronization~\cite{SalaSchoeny-Sync_TransComm,venkataramanan2010interactive,yazdi2013deterministic,ma2011efficient}. 
Due to the loss of synchronization and working over vector spaces of different dimension, correcting deletions and insertions is seen as a harder problem than correcting substitution errors.}

\ExtVe{The problem of coding for the deletion channel was introduced by Levenshtein \cite{Levenshtein-binarycodesCorrectingDeletions} in the 1960s. A set $\cC$ of binary vectors of length $n$ is a $k$-deletion correcting code if and only if any two vectors in $\cC$ do not share a common subsequence of length $n-k$. Levenshtein showed~\cite{Levenshtein-binarycodesCorrectingDeletions} that a code can correct any combination of $k$ insertions and deletions if and only if it can correct $k$ deletions. \reply{The main property of the codes being optimized is the redundancy defined as $R\triangleq n-\log|\cC|$ where $n$ is the length of the codewords in $\cC$ and $|\cC|$ is the cardinality of the code.} The optimal redundancy $n-\log |\cC|$ of any $k$-deletion correcting code $\cC$ is $\cO(k\log n)$ \cite{Levenshtein-binarycodesCorrectingDeletions}. The Varshamov-Tenengolts (VT) code~\cite{VarshTene-SingleDeletion1965} is a nearly optimal single insertion correcting code with redundancy $\log(n+1)$ bits. Constructing $k$-deletion/insertion correcting codes with small redundancy was the focus of several recent works, e.g.,~\cite{GuruswamiWang-HighNoiseHighRateDeletions_2017,brakensiek2017efficient,hanna2018guess,Gabrys-TwoDeletions_2018,Sima-TwoDeletions_2019,SimaBruck-kDeletions_2019,guruswami2020explicit}.} \reply{In~\cite{SchoenyWachterzehGabrysYaakobi-BurstDeletions-journal} and \cite{smith2017interleaved}, the authors construct codes that can correct bursts of deletions. The main idea of the papers is to imagine the codeword as a binary array and to use the structure of that array to detect and correct bursts of deletions that happen in the one-dimensional codeword.}


In this paper we extend the one-dimensional study of deletion and insertion correction to two-dimensional arrays. 
A \emph{$(\tr,\tc)$-criss-cross deletion} is the event in which an $n\times n$ array experiences a deletion of $\tr$ rows and $\tc$ columns. A code capable of correcting all $(\tr,\tc)$-criss-cross deletions is referred to as \emph{$(\tr,\tc)$-criss-cross deletion correcting code} and \emph{$(\tr,\tc)$-criss-cross insertion correcting codes} are defined similarly. 
\reply{Coding in the two dimensional space has proved profitable for data storage and wireless communications~\cite{roth1991maximum,gabidulin2008crisscrosserasure, LundGabidulinHonary-NewFamilyOptimalCorrectingTermRankErrors_2000,Sidorenko-ClassCorrectingErrorsLatticeConfiguration_1976,BlaumBruck-ArrayCodesCorrectionCrisscrossErrors_IEEE-IT2000,Gabidulin-OptimalArrayCorrectingCodes_1985,Roth-ProbabilisticCrisscrossErrorCorrection_1997,wachterzeh2017listdecodingcrisscross}. Therefore, we find it important to understand the generalization of the well-studied one-dimensional insertion- and deletion-correcting codes to the two-dimensional space. The main advantage of coding in the two-dimensional space is to leverage the structure of the code arrays rather than applying one dimensional deletion/insertion correcting codes on each dimension of the array. Along this line of thought, \cite{manabu2020delarrays} studies the problem of correcting a predetermined number of row and column deletions in two-dimensional arrays. Furthermore, the trace-reconstruction problem, which is a variant of the deletion channel, is investigated in the two-dimensional space in \cite{krishnamurthy2019trace}.}

It is well-known that in the one-dimensional case the size of the single-deletion ball equals the number of runs in the word. However, the characterization of the arrays that can be obtained from a $(1,1)$-criss-cross deletion is more complicated. 
\ExtVe{Nonetheless, we derive a non-asymptotic lower bound on the redundancy of these codes.} Second, we propose a code construction which heavily depends on the construction of non-binary single-insertion/deletion correcting codes by Tenengolts~\cite{tenengolts1984nonbinary}, \ExtVe{which can be seen as the extension of the $q$-ary alphabet of \cite{Levenshtein-binarycodesCorrectingDeletions}}. 
In the one-dimensional case, successful decoding from deletions in the transmitted word does not necessarily guarantee that the indices of the deleted symbols are known since the deletion of symbols from the same run results in the same output. While this does not impose a constraint in the one-dimensional case, we had to take this constraint into account when using non-binary single-deletion correcting codes as our component codes. 
The rest of the paper is organized as follows. In Section~\ref{sec:def}, we formally define the codes and notations that we use throughout the paper. \reply{We give a high level summary of the presented results in Section~\ref{sec:results}}. We prove in Section~\ref{sec:equiv} that the correction of $(t,t)$-criss-cross deletions and insertions is equivalent. In Section~\ref{sec:bounds}, we give a non-asymptotic upper bound on the cardinality of $(1,1)$-criss-cross deletion correcting codes. This bound shows that the minimum redundancy of these codes is $2n-3+2\log n$ bits. In Section~\ref{sec:cons}, we construct $(1,1)$-criss-cross deletion correcting codes that we call \codename codes. The correctness of this family of codes is given by an explicit decoding algorithm. 
The redundancy of the proposed \codename codes is at most $2n+4 \log n + 7 +2 \log e$.
We present in Section~\ref{sec:explicit} \codename codes with explicit encoder and decoder. We show that the explicit encoder comes at the expense of increasing the redundancy by $5\log n + 5$ bits compared to the existence result. We conclude the paper in Section~\ref{sec:concl}.

%% file: defs.tex
This section formally defines the codes and notations that we use throughout this paper. Let $\Sigma_q \triangleq \{0,\ldots, q-1\}$ be the $q$-ary alphabet. 
We denote by $\sigmatnq$ the set of all $q$-ary arrays of dimension $n \times n$. All logarithms are base $2$ unless otherwise indicated. 
For two integers $i, j \in \N$, $i \leq j$, the set $\{ i,\ldots , j \}$ is denoted by $[ i, j ]$ and the set $\{ 1,\ldots , j \}$ is denoted by $[j]$. For an array $\bfX \in \sigmatnq$,  we denote by $\xij$ the entry of $\bfX$ positioned at the $i\th$ row and $j\th$ column. We denote the $i\th$ row and $j\th$ column of $\bfX$ by $\bfX_{i,[n]}$ and $\bfX_{[n],j}$, respectively. Similarly, we denote by $\bfX_{[i_1,i_2],[j_1,j_2]}$ the sub array of $\bfX$ formed by rows $i_1$ to $i_2$ and their corresponding entries from columns $j_1$ to $j_2$.


For two positive integers $\tr, \tc \leq n$, we define a \emph{$(\tr,\tc)$-criss-cross deletion} in an array $\bfX \in \sigmatnq$ to be the deletion of any $\tr$ rows and $\tc$ columns of $\bfX$. We denote by $\dball_{\tr,\tc}(\bfX)$ the set of all arrays that result from $\bfX$ after a $(\tr,\tc)$-criss-cross deletion (i.e., the two-dimensional deletion ball\footnote{\reply{Strictly speaking, the set $\dball_{\tr,\tc}(\bfX)$ must be called the two-dimensional deletion \emph{sphere} of $\bfX$. However, we abuse terminology and refer to this set as the deletion ball to follow the nomenclature used by the literature on deletion-correcting codes. The same holds for the set $\iball_{\tr,\tc}(\bfX)$.}}). 
In a similar way we define \emph{$(\tr,\tc)$-criss-cross insertion} and the set $\iball_{\tr,\tc}(\bfX)$ for the insertion case.  If $\tr=\tc=t$, we will use the notation of $\dball_{t}(\bfX)$, \emph{$(t)$-criss-cross deletion}, \emph{$(t)$-criss-cross insertion}, and $\iball_{t}(\bfX)$.
Note that the order between the row and column deletions/insertions does not matter.

\begin{definition}[$(\tr,\tc)$-criss-cross deletion correction code] \label{def:srsc}
A  $(\tr,\tc)$-criss-cross deletion correcting code $\cC$ is a code that can correct any $(\tr,\tc)$-criss-cross deletion. 
A $(\tr,\tc)$-criss-cross insertion correcting code is defined similarly. 
\end{definition}


For clarity of presentation, we will refer to a $(\tr,\tc)$-criss-cross deletion as a \emph{$(t)$-criss-cross deletion} when $t = \tr = \tc$ and $(\tr,\tc)$-criss-cross deletion correcting code as \emph{$(t)$-criss-cross deletion correcting code}. The corresponding definitions for the insertion case are similar. Notice that throughout this paper, we do not consider combinations of insertions and deletions (c.f. Section~\ref{sec:results}).



In our code construction we use Varshamov-Tenengolts (VT) single-deletion correcting codes \cite{VarshTene-SingleDeletion1965}. A VT code was proven by Levenshtein \cite{Levenshtein-binarycodesCorrectingDeletions} to correct a single deletion in a binary string of length $n$, with redundancy not more than $\log(n+1)$ bits. In fact, we use Tenengolt's extension~\cite{tenengolts1984nonbinary} for the $q$-ary alphabet, which is briefly explained next. For a $q$-ary vector $\mathbf{x}=(x_1,\dots, x_n)$ we associate its \emph{binary signature} $\bfs=(s_1,\dots,s_{n})$. The entries of $\mathbf{s}$ are calculated such that $s_1 = 1$ and $s_i = 1$ if $x_i \geq x_{i-1}$ or $s_i = 0$ otherwise for $i>1$. Thus, all $q$-ary vectors of length $n$ can be split into disjoint cosets $\VTab{q}$ defined as the set of all $\mathbf{x}$ with signature $\mathbf{s}$ satisfying
\begin{align*}
\sum_{i=1}^{n}(i-1) s_i \equiv a \mod n, \quad & \sum_{i=1}^n x_i \equiv b \mod q,
\end{align*}
where $0\leq a\leq n-1,0\leq b\leq q-1$. Each coset is a single $q$-ary insertion/deletion correcting code. 
Note that the $qn$ disjoint cosets form a partition of $\Sigma_q^n$. Therefore, by the pigeon-hole principal, there exists a set (or a VT code)  $\cV\cT_{n,q}(a^\star,b^\star)$ such that \begin{equation*}
    |\cV\cT_{n,q}(a^\star,b^\star)| \geq \frac{q^n}{qn}.
\end{equation*}

%% file: results.tex
Our main results can be summarized as follows. In Theorem~\ref{theorem:equiv}, we extend the equivalence between insertion correcting codes and deletion correcting to the $2$-dimensional codes considered in this setting. Namely, we show that a given code $\mathcal{C}$ is a $(t)$-criss-cross deletion correcting code if and only if $\mathcal{C}$ is a $(t)$-criss-cross insertion correcting code. As a consequence, all our results proven for the \crisscross deletion case hold for the insertion case as well.
%
%
%
To evaluate how good a given \crisscross deletion correcting code is, we derive a non-asymptotic upper bound on the cardinality of any criss-cross deletion correcting code as follows.

\emph{Lower bound:} (Theorem~\ref{thm:non-asymp-upper_bound})
The non-asymptotic redundancy of a $q$-ary $(1)$-criss-cross deletion correcting code $\cC$ is bounded from below by
$R \geq 2n -3 +2\log_q n.$

We show that there exist $(1)$-criss-cross deletion/insertion correction codes that have redundancy $2\log n+o(1)$ bits far from our lower bound. We do so by constructing an existential \crisscross deletion correction code called \codename code that has redundancy $2n + 4\log n + o(1)$. We extend the existential construction to a \crisscross code with explicit encoder and decoder at the expense of increasing the redundancy by $5\log n + 5$ bits.

\emph{Code constructions:} The \codename code constructed in Section~\ref{sec:cons} is a \crisscross deletion/insertion correcting code (Theorem~\ref{thm:cons}). The redundancy of the code is upper bounded by $2n+4\log n +o(1)$ bits (Corollary~\ref{corr:red}). The encoder of the code can be made systematic at the expense of increasing the redundancy to at most $2n+ 9\log n + o(1)$ bits (Theorem~\ref{thm:exp_cons}).

\reply{\emph{Extensions:} In this work we restrict our attention to $(t)$-criss-cross deletions and insertions. However, our $(t)$-criss-cross deletion-correcting code construction can correct a mixed $(1)$-criss-cross error defined as a row/column deletion and a column/row insertion. In addition our codes can correct a single row insertion/deletion or a single column insertion/deletion. Nevertheless, the bound on the redundancy does not necessarily hold for those general problems. In fact, we show in~\cite{welter2021multiple} that, under the generalized model, a code correcting a $(1)$-criss-cross deletion is able to correct two row deletions (no column deletions) or two column deletions (no row deletions). 
We leave those general problems as an interesting direction for future research.}

%% file: equivalence.tex
\ExtVe{In this section we first show an equivalence between a $(1)$-criss-cross deletion correcting code and a
$(1)$-criss-cross insertion correcting code (Theorem~\ref{theorem:equiv}). Then we use the result of Theorem~\ref{theorem:equiv} to prove the more general equivalence between ($t$)-criss-cross deletion correcting codes and ($t$)-criss-cross insertion correcting codes for all $t\in [n-1]$ (Corollary~\ref{cor:tinsdel})}.
\begin{theorem} \label{theorem:equiv}
A code $\cC\subset \sigmatnq$ is a $(1)$-criss-cross deletion correcting code if and only if it is a 
$(1)$-criss-cross insertion correcting code.
\end{theorem}

 \begin{cor}\label{cor:tinsdel}
 For any integer $t\in [n-1]$, a code $\cC\subset\oi ^{n\times n} _q$ is a ($t$)-criss-cross deletion correcting code if and only if it is a ($t$)-criss-cross insertion correcting code.
 \end{cor}

Note that in the one-dimensional case Theorem~\ref{theorem:equiv} holds since the intersection of the deletion balls of two vectors is not trivial if and only if the intersection of their insertion balls is not trivial \cite{Levenshtein-binarycodesCorrectingDeletions}. Since this property holds over any alphabet, the following lemma can be derived by considering the arrays as one dimensional vectors where each element is a row/column. 
 \begin{lemma} \label{lemma:equivsimpl}
For a positive integer $m$ and two arrays $\bfX\in \oi^{m\times m}_q,\bfY\in \oi^{m\times m}_q$, 
\begin{align*}
    &\dball_{1,0}(\bfX)\cap\dball_{1,0}(\bfY)\neq\emptyset \textrm{ if and only if } \iball_{1,0}(\bfX)\cap\iball_{1,0}(\bfY)\neq\emptyset \\
    &\dball_{0,1}(\bfX)\cap\dball_{0,1}(\bfY)\neq\emptyset \textrm{ if and only if } \iball_{0,1}(\bfX)\cap\iball_{0,1}(\bfY)\neq\emptyset.
\end{align*}
 \end{lemma}

While the last lemma is derived from properties of vectors, the next one, albeit similar, requires a complete proof. 
\begin{lemma} \label{lemma:equiv}
For a positive integer $m$ and two arrays $\bfX\in \oi^{(m+1)\times m}_q,\bfY\in \oi^{m\times (m+1)}_q$,
$$\dball_{1,0}(\bfX)\cap\dball_{0,1}(\bfY)\neq\emptyset \textrm{ if and only if } \iball_{0,1}(\bfX)\cap\iball_{1,0}(\bfY)\neq\emptyset.$$
\end{lemma}

\begin{IEEEproof}
We show the ``if" direction while the ``only if" part is proved similarly. That is, we prove that if $\dball_{1,0}(\bfX)\cap\dball_{0,1}(\bfY)\neq\emptyset$ then 
$\iball_{0,1}(\bfX)\cap\iball_{1,0}(\bfY)\neq\emptyset$.  Assume that there exists $\bfD\in\oi^{m\times m}_q$ such that $\bfD\in\dball_{1,0}(\bfX)\cap\dball_{0,1}(\bfY)$ and by contradiction assume that $\iball_{0,1}(\bfX)\cap\iball_{1,0}(\bfY) = \emptyset$. Let 
$\rIndex,\cIndex$ be the indices of the row and column deleted in $\bfX$ and $\bfY$, respectively, to obtain $\bfD$. 
Let $\bfr$ denote row $\rIndex$ of $\bfX$, i.e., $\bfX_{\rIndex,[m]}$, after an insertion of 0 in position $\cIndex$. Similarly, let $\bfc$ be the column $\bfY_{[m],\cIndex}$ after an insertion of 0 in position $\rIndex$. Notice that it is also possible to insert $1$ in both of the words, as long as the symbol inserted is the same. The following relations hold from the definition of $\bfD$.
\begin{equation}\label{eq:equivalence}
 \begin{split}
 &\xij=\dij=\yij\textrm{ for }1\leq i<\rIndex,1\leq j<\cIndex,\\
 &X_{i+1,j}=\dij=\yij\textrm{ for } \rIndex\leq i\leq m,1\leq j<\cIndex,\\
 &\xij=\dij=Y_{i,j+1}\textrm{ for } 1\leq i<\rIndex,\cIndex\leq j\leq m,\\
 &X_{i+1,j}=\dij=Y_{i,j+1}\textrm{ for }\rIndex\leq i \leq m, \cIndex\leq j \leq m.
 \end{split}
 \end{equation}
Let $\iX$ be the result of inserting column $\bfc$ at index $\cIndex$ into $\bfX$. The array $\iY$ is defined similarly by inserting row $\bfr$ at index $\rIndex$ in $\bfY$. Notice that $\iX$ is a result of inserting a column to $\bfX$ and thus $\iX\in \iball_{0,1}(\bfX)$. For the same reasons it holds that $\iY\in \iball_{1,0}(\bfY)$.
We conclude the proof by showing that $\iX=\iY$. This will be done by considering the following cases.
\begin{description}[font=$\bullet$\scshape\bfseries]
\item For $i<\rIndex, j<\cIndex$, both $\ixij,\iyij$ are not affected by the insertions or deletions. Hence, it follows that
$$\ixij=\xij=\yij=\iyij$$
\item For $i=\rIndex$ and for $j<\cIndex$, the symbols $\ixij=r_j$ remain unaffected by the insertion.
On the other hand, $\iyij$ is exactly an inserted symbol into $\bfY$ that is defined to be $r_j$ which results in $\iyij=r_j=\ixij.$
\item For $i<\rIndex$ and for $j=\cIndex$, the symbols $\iyij=c_i$ remain unaffected by the insertion.
On the other hand, $\ixij$ is exactly an inserted symbol into $\bfX$ that is defined to be $c_i$ which results in $\ixij=c_i=\iyij.$
\item For $i=\rIndex$ and for $j=\cIndex$, it holds that $\ixij=c_i$ and $\iyij=r_j$. By definition, both of these symbols are 0, which results in $\ixij=\iyij.$
\item For $i>\rIndex$ and for $j<\cIndex$, by~\eqref{eq:equivalence} it holds that 
$$\ixij =\xij= D_{i-1,j}=Y_{i-1,j}.$$
On the other hand, after a row insertion in index $\rIndex$, it holds that $\iyij=Y_{i-1,j}$ which results in $\ixij=\iyij.$
\item For $i>\rIndex$ and for $j=\cIndex$, by definition $\ixij=c_i=Y_{i-1,j}$. On the other hand, $\iY$ had a row insertion in index $\rIndex$, which means that $\iyij=Y_{i-1,j}$ and results in $\ixij=\iyij.$
\item For $i<\rIndex$ and for $j>\cIndex$, by~\eqref{eq:equivalence} it holds that 
$$\iyij =\yij= D_{i,j-1}=X_{i,j-1}.$$
On the other hand, after a column insertion in index $\cIndex$, it holds that $\ixij=X_{i,j-1}$ which results in $\ixij=\iyij.$
\item For $i=\rIndex$ and for $j>\cIndex$, by definition $\iyij=r_j=X_{i,j-1}$. On the other hand, $\iX$ had a column insertion in index $\cIndex$, which means that $\ixij=X_{i,j-1}$ and results in $\ixij=\iyij.$
\item For $i>\rIndex$ and for $j>\cIndex$, $\iX$ had a column insertion in index $\cIndex$, which means that $\ixij=X_{i,j-1}$. On the other hand, $\iY$ had a row insertion in index $\rIndex$, which means that $\iyij=Y_{i-1,j}$. From~\eqref{eq:equivalence} it holds that $X_{i,j-1}=D_{i-1,j-1}$ and $Y_{i-1,j}=D_{i-1,j-1}$. This results in 
$$\ixij=X_{i,j-1} = D_{i-1,j-1} = Y_{i-1,j} = \iyij.$$
\end{description}
This concludes that for all $i,j\in [m+1]$, $\ixij=\iyij$, which assures that $\iX =  \iY$, and hence  $\iball_{0,1}(\bfX)\cap\iball_{1,0}(\bfY) \neq\emptyset$, that contradicts our assumption. 
 \end{IEEEproof}




We now use the results of  Lemma~\ref{lemma:equivsimpl} and Lemma~\ref{lemma:equiv} to prove Theorem~\ref{theorem:equiv}.
 \begin{IEEEproof}[Proof of Theorem~\ref{theorem:equiv}]
 \begin{figure}
 \centering
     \resizebox{.5\totalheight}{!}{\input{EquivGraph.tex}}
     \caption{ A flowchart of the proof of Theorem~\ref{theorem:equiv}. } 
     \label{fig:equivGraph}
 \end{figure}
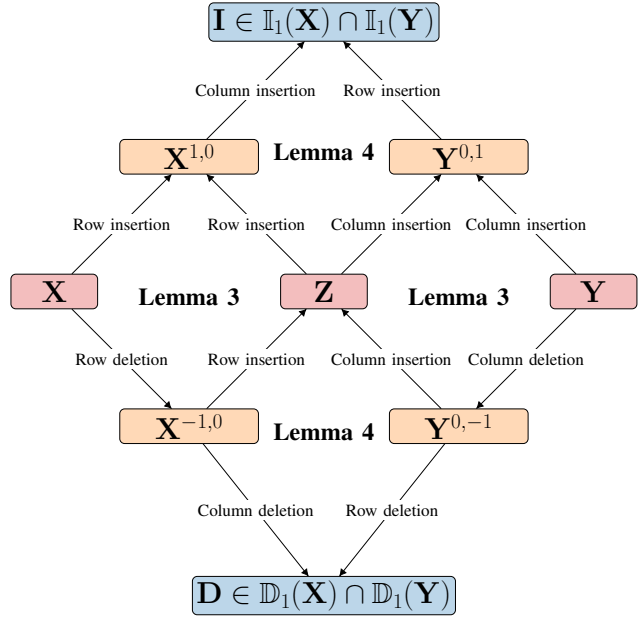
The proof follows by showing that for any $\bfX,\bfY\in \sigmatnq$, $\dball_{1}(\bfX)\cap \dball_{1}(\bfY)=\emptyset$ if and only if $\iball_{1}(\bfX)\cap \iball_{1}(\bfY)=\emptyset$. For the reader's convenience, a flowchart of the proof is presented in Figure~\ref{fig:equivGraph}. We only show the ``only if" part as the ``if" part follows similarly. 

Assume that there exists an array $\bfD \in \oi^{(n-1)\times(n-1)}_q$ such that $\bfD \in \dball_{1,1}(\bfX)\cap\dball_{1,1}(\bfY)$. Hence, $\bfD$ can be obtained by deleting a row and then a column from $\bfX$ and by deleting a column and then a row from $\bfY$ (note that the order of the row and column deletions does not matter and can be chosen arbitrarily). 
 Denote the intermediate arrays by $\xMidDel,\yMidDel$, so the following relation holds. 
 \begin{align*}
     \bfX \xrightarrow{\textrm{Row Deletion}} &\xMidDel \xrightarrow{\textrm{Col Deletion}} \bfD, \\
     \bfY \xrightarrow{\textrm{Col Deletion}} &\yMidDel \xrightarrow{\textrm{Row Deletion}} \bfD.
 \end{align*}
 Hence, it holds that $$\bfD\in \dball_{1,0}(\yMidDel)\cap\dball_{0,1}(\xMidDel),$$
 and thus, from Lemma~\ref{lemma:equiv} there exists an array $\bfZ\in \sigmatnq$, such that  $\bfZ\in\iball_{0,1}(\yMidDel)\cap\iball_{1,0}(\xMidDel)$.
 By definition, $\bfZ\in \iball_{1,0}(\xMidDel)$ is equivalent to 
 $\xMidDel \in \dball_{1,0}(\bfZ)$. But, it is also known that $\xMidDel \in \dball_{1,0}(\bfX)$, which means that 
 $$\xMidDel \in \dball_{1,0}(\bfZ)\cap \dball_{1,0}(\bfX).$$
 From Lemma~\ref{lemma:equivsimpl} it follows that there exists some $\xMidIns\in \iball_{1,0}(\bfZ)\cap \iball_{1,0}(\bfX)$. The same argument can be done for $\bfY$, and define its result by  $\yMidIns$. Next, notice that we can also conclude that 
 $$\bfZ\in \dball_{1,0}(\xMidIns)\cap \dball_{0,1}(\yMidIns),$$
so from Lemma~\ref{lemma:equiv} it is deduced that there exists an array $\bfI\in\iball_{0,1}(\xMidIns)\cap \iball_{1,0}(\yMidIns)$.
 Note that $\bfI\in \iball_{0,1}(\xMidIns)$ and $\xMidIns\in~\iball_{1,0}(\bfX)$, which means $\bfI$ is obtained by inserting a row and a column in $\bfX$, i.e., $\bfI\in \iball_{1}(\bfX)$. A symmetrical argument holds for $\bfY$, which assures that $\bfI\in \iball_{1}(\bfX)\cap \iball_{1}(\bfY).$
 \end{IEEEproof}
 
We now prove Corollary~\ref{cor:tinsdel} by using the result of Theorem~\ref{theorem:equiv}.

 \begin{IEEEproof}[Proof of Corollary~\ref{cor:tinsdel}]
The proof follows by showing that for any $\bfX_1,\bfX_{t+1}\in \sigmatnq$, $\dball_{t}(\bfX)\cap \dball_{t}(\bfX_{t+1})\neq \emptyset$ if and only if $\iball_{t}(\bfX)\cap \iball_{t}(\bfX_{t+1}) \neq \emptyset$. We first prove the following claim.

\begin{claim}\label{claim:deletion}
 For any two arrays $\bfX_1, \bfX_{t+1}\in \sigmatnq$, $\dball_{t}(\bfX_1)\cap \dball_{t}(\bfX_{t+1})\neq \emptyset$ if and only if there exist $t-1$ arrays $\bfX_2,\dots,\bfX_{t}$ such that $\dball_{1}(\bfX_i)\cap \dball_{1}(\bfX_{i+1})\neq \emptyset$ for all $1\leq i \leq t$.
\end{claim}

\begin{IEEEproof}
 We prove the ``if'' part by induction. The proof of the ``only if'' part follows similarly and is omitted.
 
 \paragraph{Base case} We need to show that if $\dball_{1}(\bfX_1)\cap \dball_{1}(\bfX_{2})\neq \emptyset$ then $\dball_{1}(\bfX_i)\cap \dball_{1}(\bfX_{i+1})\neq \emptyset$ for all $i=1$ which follows from the assumption.
 
\paragraph{Induction step}
Assume the property holds for $t\in [n-2]$ and we show that the property holds for $t+1$. Let $\bfX_1, \bfX_{t+2}$ be such that $\dball_{t+1}(\bfX_1)\cap \dball_{t+1}(\bfX_{t+2})\neq \emptyset$. Then, there exists $\bfXi{1}_1,\bfXi{1}_{t+1}$ resulting from a criss-cross deletion of $\bfX_1$ and $\bfX_{t+2}$, respectively, such that $\dball_{t}(\bfXi{1}_1\cap \dball_{t}(\bfXi{1}_{t+1})\neq \emptyset$. Thus, according to the induction hypothesis, there exist $t-2$ arrays $\bfXi{1}_1,\dots,\bfXi{1}_{t}$ that satisfy $\dball_{1}(\bfXi{1}_i)\cap \dball_{1}(\bfXi{1}_{i+1})\neq \emptyset$ for all $1\leq i \leq t$.

\noindent According to Theorem~\ref{theorem:equiv}, there exist $t$ arrays $\bfX_2,\dots,\bfX_{t+1}$ such that for all $2\leq i \leq t+1$, $\bfX_i \in \iball(\bfXi{1}_{i-1})\cap \iball(\bfXi{1}_{i})$. Therefore, it holds that for $1\leq 1 \leq t+1$,
\begin{equation*}
    \bfXi{1}_i \in \dball_1(\bfX_{i})\cap \dball_1(\bfX_{i+1}).
\end{equation*}
This completes the ``if'' part of the proof.
\end{IEEEproof}

Next we prove a similar claim for the insertion case.

\begin{claim}\label{claim:insertion}
 For any two arrays $\bfX_1, \bfX_{t+1}\in \sigmatnq$, $\iball_{t}(\bfX_1)\cap \iball_{t}(\bfX_{t+1})\neq \emptyset$ if and only if there exist $t-1$ arrays $\bfX_2,\dots,\bfX_{t}$ such that $\iball_{1}(\bfX_i)\cap \iball_{1}(\bfX_{i+1})\neq \emptyset$ for all $1\leq i \leq t$.
\end{claim}
The proof of Claim~\ref{claim:insertion} is similar to the proof of Claim~\ref{claim:deletion} and is thus given in Appendix~\ref{app:claim_ins}.

Having the results of Claim~\ref{claim:deletion} and Claim~\ref{claim:insertion}, we can now prove Corollary~\ref{cor:tinsdel} as follows. For any $\bfX_t,\bfX_{t+1} \in \sigmatnq$, if $\dball_{t}(\bfX_1)\cap \dball_{t}(\bfX_{t+1})\neq \emptyset$, then from Claim~\ref{claim:deletion} we know that there exist $t-1$ arrays $\bfX_2,\dots,\bfX_t$ such that $\dball_{1}(\bfX_i)\cap \dball_{1}(\bfX_{i+1})\neq \emptyset$ for all $1\leq i \leq t$.
Then, according to Theorem~\ref{theorem:equiv}, there exist $t$ arrays $\bfXi{1}_1,\dots,\bfXi{1}_{t}$ such that for all $1\leq i \leq t$,
\begin{equation*}
    \bfXi{1}_i \in \iball_1(\bfX_{i})\cap \iball_1(\bfX_{i+1}).
\end{equation*}
Finally, we can now apply Claim~\ref{claim:insertion} to conclude that $\iball_{t}(\bfX_1)\cap \iball_{t}(\bfX_{t+1})\neq \emptyset$. The ``only if'' part follows similarly.
\end{IEEEproof}

%% file: EquivGraph.tex
\definecolor{color0}{rgb}{0.12156862745098,0.466666666666667,0.705882352941177}
\definecolor{color1}{rgb}{1,0.498039215686275,0.0549019607843137}
\definecolor{color2}{rgb}{0.172549019607843,0.627450980392157,0.172549019607843}
\definecolor{color3}{rgb}{0.83921568627451,0.152941176470588,0.156862745098039}
\definecolor{color4}{rgb}{0.580392156862745,0.403921568627451,0.741176470588235}
\definecolor{color6}{rgb}{0.549019607843137,0.337254901960784,0.294117647058824}
\definecolor{color5}{rgb}{0.890196078431372,0.466666666666667,0.76078431372549}

\tikzset{%
  >={Latex[width=2mm,length=2mm]},
            base/.style = {rectangle, rounded corners, draw=black,
                           minimum width=4cm, minimum height=1cm,
                           text centered, font=\Huge},
  activityStarts/.style = {base, fill=color0!30},
    activityRuns/.style = {base, fill=color1!30},
         process/.style = {base, minimum width=2.5cm, fill=color3!30,
                           font=\Huge},
         lemma/.style = { minimum width=2.5cm, ,
                           font=\huge},
}

\begin{tikzpicture}[node distance=5.5cm, 
    every node/.style={fill=white, font=\Large }, align=center]
  \node (start)             [activityStarts]              {$\bfI\in \iball_{1}(\bfX)\cap                                                                        \iball_{1}(\bfY)$};
  \node (XinsBlock)     [activityRuns, below left of= start]          {$\xMidIns$};
  \node (YinsBlock)     [activityRuns, below right of= start]          {$\yMidIns$};
  \node (LemmaSimplblock)      [ lemma, below of=XinsBlock, yshift=14.6mm]                                                                                   {\textbf{Lemma~\ref{lemma:equivsimpl}}};
  \node (LemmaSimplblock2)      [ lemma, below of=YinsBlock, yshift=14.6mm]                                                                                 {\textbf{Lemma~\ref{lemma:equivsimpl}}};

  \node (Xblock)      [process, below left of=XinsBlock]   {$\bfX$};
  \node (Zblock)      [process, below right of=XinsBlock]   {$\bfZ$};
  \node (Lemmablock)      [ lemma, below of=Zblock, yshift=14.6mm]   {\textbf{Lemma~\ref{lemma:equiv}}};
  \node (Lemmablock2)      [ lemma, above of=Zblock, yshift=-14.6mm]   {\textbf{Lemma~\ref{lemma:equiv}}};
  \node (Yblock)      [process, below right of=YinsBlock]   {$\bfY$};
  \node (XdelBlock)      [activityRuns, below right of=Xblock]
                                                      {$\xMidDel$};
  \node (YdelBlock)      [activityRuns, below right of=Zblock]
                                                      {$\yMidDel$};
  \node (CommonDelBlock)      [activityStarts, below right of=XdelBlock, yshift=-1cm]
        {$\bfD\in \dball_{1}(\bfX)\cap \dball_{1}(\bfY)$};
  \draw[->]     (XinsBlock) -- node[text width=4cm]
                                   {Column~insertion}(start);
  \draw[->]     (YinsBlock) -- node[text width=4cm]
                                   {Row~insertion} (start);
  \draw[->]     (Xblock) -- node[text width=4cm]
                                   {Row~insertion} (XinsBlock);
  \draw[->]     (Zblock) -- node[text width=4cm]
                                   {Column~insertion}(YinsBlock);
  \draw[->]     (Yblock) -- node[text width=4cm]
                                   {Column~insertion}(YinsBlock);
  \draw[->]     (Zblock) -- node[text width=4cm]
                                   {Row~insertion} (XinsBlock);
  \draw[->]      (Xblock) -- node[text width=4cm]
                                   {Row~deletion} (XdelBlock);
  \draw[->]      (XdelBlock) --  node[text width=4cm]
                                   {Row~insertion} (Zblock);
  \draw[->]      (YdelBlock) --  node[text width=4cm]
                                   {Column~insertion} (Zblock);
  \draw[->]      (Yblock) -- node[text width=4cm]
                                   {Column~deletion} (YdelBlock);

  \draw[->]      (XdelBlock) -- node[text width=3.5cm]
                                   {Column~deletion} (CommonDelBlock);
  \draw[->]      (YdelBlock) -- node[text width=4cm]
                                   {Row~deletion} (CommonDelBlock);

\end{tikzpicture}

%% file: bounds.tex
%

\begin{figure}
	\centering
	\resizebox{.9\totalheight}{!}{
	\input{bad_matrix.tikz}
	}
	\caption{Required pattern for $\bfX^{i_1,j_1} = \bfX^{i_2,j_2}$. Let $(i_1,j_1)\neq(i_2,j_2)$ be the indices of the deleted row and column in two different criss-cross deletions on the array $\bfX$. W.l.o.g $j_1<j_2$ and for case $i_1<i_2$ the constraints are: \textit{1)} Each row of the sub arrays $\Xt{}$ and $\Xb{}$ must be a row run of length $j_2-j_1+1$ and each column of $\Xl{}$ and $\Xr{}$ must be a column run of length $i_2-i_1+1$. 
	\textit{2)} Each diagonal of the sub array $\Xc{}$ must be a diagonal run.
    \textit{3)} The corner sub arrays $\XI{}$, $\Xii{}$, $\Xiii{}$ and $\Xiv{}$ are outside of the region affected by the criss-cross deletions. Therefore, no constraints are imposed on those sub arrays. The same holds for $X_{i_1,j_2}$ and $X_{i_2,j_1}$ since they are both deleted by criss-cross deletions.
Note that for $i_1 > i_2$, all the requirements remain the same except for $\Xc{}$. In this case, the bottom-left to top-right diagonals are diagonal runs.} 
	\label{fig:badmatrix}
\end{figure}
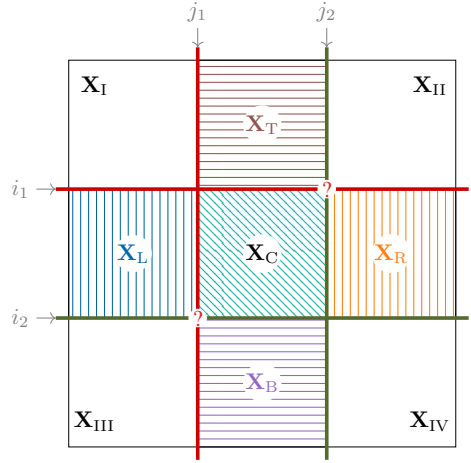

In this section we prove a non-asymptotic upper bound on the cardinality of a \crisscross deletion correcting code. For an array $\bfX\in \sigmatnq$, we denote by $\bfX^{i,j}$ the array obtained from $\bfX$ after deleting the $i\th$ row and the $j\th$ column. Let $\bfX \in \sigmatnq$ and let $i_1,i_2,j_1 \in [n]$ be such that $i_1\leq i_2$. We define a \emph{column run} of length $i_2-i_1+1$ as a sequence of identical consecutive symbols in a column $j_1$, i.e., $X_{i_1,j_1} = X_{i_1+1,j_1} = \cdots = X_{i_2,j_1}$. We define a \emph{row run} similarly. A \emph{diagonal run} of length $\delta + 1$ is a sequence of identical symbols situated on a diagonal of $\bfX$, i.e., $X_{i_1,j_1} = X_{i_1+1,j_1+1} = \cdots = X_{i_1+\delta,j_1+\delta}$. \reply{An \emph{anti-diagonal run} of length $\delta +1$ is a sequence of identical symbols situated on an anti-diagonal of $\bfX$, i.e., $X_{i_1,j_1} = X_{i_1+1,j_1-1} = \cdots = X_{i_1+\delta,j_1-\delta}$.}
In Lemma~\ref{lem:del-pattern} we give a necessary and sufficient condition that two different \crisscross deletions applied on an array $\bfX$ must satisfy to result in the same array $\bfX^{i_1,j_1} = \bfX^{i_2,j_2}$ for $(i_1,j_1)\neq (i_2,j_2)$. \reply{We start with an example that illustrates the idea of Lemma~\ref{lem:del-pattern}.}
\begin{example}
\reply{Consider the following binary $9\times 9$ array divided into nine $3\times 3$ arrays structured as in Figure~\ref{fig:badmatrix}.}
\definecolor{color0}{rgb}{0.12156862745098,0.466666666666667,0.705882352941177}
\definecolor{color1}{rgb}{1,0.498039215686275,0.0549019607843137}
\definecolor{color2}{rgb}{0.172549019607843,0.627450980392157,0.172549019607843}
\definecolor{color3}{rgb}{0.83921568627451,0.152941176470588,0.156862745098039}
\definecolor{color4}{rgb}{0.580392156862745,0.403921568627451,0.741176470588235}
\definecolor{color6}{rgb}{0.549019607843137,0.337254901960784,0.294117647058824}
\definecolor{color5}{rgb}{0.890196078431372,0.466666666666667,0.76078431372549}

\definecolor{lightseagreen}{rgb}{0.13, 0.7, 0.67}
\definecolor{darkolivegreen}{rgb}{0.33, 0.42, 0.18}
\definecolor{mediumpersianblue}{rgb}{0.0, 0.4, 0.65}
\definecolor{bostonuniversityred}{rgb}{0.8, 0.0, 0.0}
\begin{equation*}
   \bfX = \begin{bmatrix}
   \XI{} & \color{color6} \XT & \Xii{}\\
   \color{mediumpersianblue} \XL & \color{lightseagreen} \XC & \color{color1} \XR\\
   \Xiii{} & \color{color4} \XB & \Xiv{}\\
   \end{bmatrix} = \begin{bmatrix}
        0 & 1 & 0 & \color{color6}1 & \color{color6} 1 & \color{color6} 1 & 0 & 1 & 0\\
        1 & 1 & 1 & \color{color6}0 & \color{color6} 0 & \color{color6} 0 & 1 & 0 & 0\\
        1 & 0 & 1 & \color{color6}0 & \color{color6} 0 & \color{color6} 0 & 0 & 1 & 0\\
        \color{mediumpersianblue} 0 & \color{mediumpersianblue} 1 & \color{mediumpersianblue} 1 & \color{lightseagreen} 1 & \color{lightseagreen} 0 & \color{lightseagreen} 1 & \color{color1} 0 & \color{color1} 1 & \color{color1} 0\\
        \color{mediumpersianblue} 0 & \color{mediumpersianblue} 1 & \color{mediumpersianblue} 1 & \color{lightseagreen} 0 & \color{lightseagreen} 1 & \color{lightseagreen} 0 & \color{color1} 0 & \color{color1} 1 & \color{color1} 0\\
        \color{mediumpersianblue} 0 & \color{mediumpersianblue} 1 & \color{mediumpersianblue} 1 & \color{lightseagreen} 0 & \color{lightseagreen} 0 & \color{lightseagreen} 1 & \color{color1} 0 & \color{color1} 1 & \color{color1} 0\\
        0 & 1 & 0 & \color{color4} 0 & \color{color4} 0 & \color{color4} 0 & 0 & 1 & 0\\
        1 & 1 & 1 & \color{color4} 1 & \color{color4} 1 & \color{color4} 1 & 1 & 0 & 0\\
        1 & 0 & 1 & \color{color4} 1 & \color{color4} 1 & \color{color4} 1 & 0 & 1 & 0\\
    \end{bmatrix}.
\end{equation*}
\reply{It is easy to verify that deleting column $4$ and row $4$ or deleting column $6$ and row $6$ results in the same array, i.e., $\bfX^{4,4} = \bfX^{6,6}$. Let $(i_1,j_1) = (4,4)$ and $(i_2,j_2) = (6,6)$. The equality $\bfX^{4,4} = \bfX^{6,6}$ happens because: all rows of the arrays $\XT = \bfX_{[1,i_1-1],[j_1,j_2]} = \bfX_{[1,3],[4,6]}$ and $\XB= \bfX_{[i_2+1,n],[j_1,j_2]} = \bfX_{[7,9],[4,6]}$ are row runs; all the columns of the arrays $\XL = \bfX_{[i_1,i_2],[1,j_1-1]} = \bfX_{[4,6],[1,3]}$ and $\XR = \bfX_{[i_1,i_2],[j_2+1,n]} = \bfX_{[4,6],[7,9]}$ are column runs; and all the diagonals of $\XC = \bfX_{[i_1,i_2],[j_1,j_2]}$ are diagonal runs. Lemma~\ref{lem:del-pattern} generalises this example to show that given an array $\bfX$ and two (1)-criss-cross deletions applied on $\bfX$, the equality $\bfX^{i_1,j_1} = \bfX^{i_2,j_2}$ for $(i_1,j_1)\neq (i_2,j_2)$ holds if and only if $\bfX$ has the structure described in this example.}
\end{example}

\begin{lem} \label{lem:del-pattern}
For $i_1,i_2,j_1,j_2 \in [n]$ such that $(i_1,j_1)\neq(i_2,j_2)$, we define $\imin \triangleq \text{\em min}(i_1,i_2)$ and $\imax \triangleq \text{\em max}(i_1,i_2)$ and assume w.l.o.g. that $j_1 \leq j_2$. For all $n\geq 3$ and $\bfX \in \sigmatnq$, the equality $\bfX^{i_1,j_1} = \bfX^{i_2,j_2}$ holds if and only if the entries $X_{i,j}$ of $\bfX$ satisfy the following structure (illustrated in Figure~\ref{fig:badmatrix} for the case $i_1 \leq i_2$).
\begin{table}[h!]
\centering
\setlength{\tabcolsep}{.01pt}
\begin{tabular}{l c}
     $i \in [1,\imin-1], j\in [1,j_1-1]$: & $X_{i,j}$ is arbitrary, \\
     $i \in [1,\imin-1], j \in [j_2+1,n]$: & $X_{i,j}$ is arbitrary, \\
     $i \in [\imax+1,n], j \in [1,j_1-1]$: & $X_{i,j}$ is arbitrary, \\
     $i \in [\imax+1,n], j \in [j_2+1,n]$: & $X_{i,j}$ is arbitrary, \\
     $i\in [1,\imin], j \in [j_1,j_2-1]$: & $X_{i,j} = X_{i,j+1}$, \\
     $i \in [\imax+1,n],j \in [j_1,j_2-1]$: & $X_{i,j} = X_{i,j+1}$, \\
     $i \in [\imin,\imax-1],j \in [1,j_1-1]$: & $X_{i,j} = X_{i+1,j}$,  \\
     $i \in [\imin,\imax-1],j \in [j_2+1,n]$: & $X_{i,j} = X_{i+1,j}$,  \\
$\begin{cases} 
i \in [\imin,\imax-1],j \in [j_1+1,j_2]:\\
i \in [\imin,\imax-1],j \in [j_1,j_2-1]:
\end{cases}$
& $
\begin{aligned}
X_{i,j} = X_{i+1,j+1} & \text{ for } i_1\leq i_2,\\
X_{i,j} = X_{i+1,j-1} & \text{ for } i_1>i_2.
\end{aligned}$
\end{tabular}
\end{table}
\end{lem}
%
%

\begin{IEEEproof}
For a better grasp of the proof we use the notation of Figure~\ref{fig:badmatrix} for the sub arrays and the aforementioned notation of runs. Furthermore, for an array $\bfX$ we write $\bfX_{[:-1]}$ if all the elements of that array are shifted by one column to the left, \ie $\bfX_{[:-1]} = \left[\bfX_{[n],2}|\bfX_{[n],3}|\cdots|\bfX_{[n],n}\right]$ where $[\cdot|\cdot]$ denotes a concatenation of arrays. Similarly, we write $\bfX_{[-1:]}$ for a row shift by one to the top and $\bfX_{[-1:-1]}$ for a simultaneous row and column shift.

We now assume that for $(i_1,j_1)\neq (i_2,j_2) \in [n]\times [n]$ there exists an array $\widetilde{\bfX} \in \sigmatniq$ such that $\widetilde{\bfX} = \bfX^{i_1,j_1} = \bfX^{i_2,j_2}$. Note that w.l.o.g. we assume that $j_1 \leq j_2$.

Due to the assumption that $\bfX^{i_1,j_1} = \bfX^{i_2,j_2}$ we have  $\XI{}^{i_1,j_1}  = \XI{}^{i_2,j_2}$. \reply{The indices of the columns and rows of $\XI{}^{i_1,j_1}$  and $\XI{}^{i_2,j_2}$ satisfy $i < \imin$ and $j < j_1$. Therefore the entries of those sub arrays are not affected by the criss-cross deletion. Hence, $\XI{}^{i_1,j_1} = \XI{}^{i_2,j_2} =\XI{}$ irrespective of the values of the entries of $\bfX$.} 
For $i < \imin$ and $j > j_2$, we have $\Xii{}^{i_1,j_1} = \Xii{}^{i_2,j_2} = \Xii{[:-1]}$ irrespective of the values of the entries of $\bfX$. This equality holds because in both cases the columns of $\Xii{}$ are shifted by one column to the left. Using a similar argument, one can show that $\Xiii{}^{i_1,j_1} = \Xiii{}^{i_2,j_2} = \Xiii{[-1:]}$ and $\Xiv{}^{i_1,j_1} = \Xiv{}^{i_2,j_2} = \Xiv{[-1:-1]}$.


In contrast, from $\Xt{}^{i_1,j_1}  = \Xt{}^{i_2,j_2}$ we get $\Xt{}^{i_1,j_1}=\Xt {[:-1]}$ and $\Xt{}^{i_2,j_2} = \Xt{}$ which produces the row run constraint $\Xt {[:-1]}=\Xt{}$, i.e., $X_{i,j} = X_{i,j+1}$ for all corresponding values of $i$ and $j$. The same constraints hold for the equalities $\Xb{}^{i_1,j_1}=\Xb {[-1:-1]}$ and $\Xb{}^{i_2,j_2} =\Xb {[-1:]}$ following from the existence of $\widetilde{\bfX}$.

Furthermore, we observe that $\Xl{}^{i_1,j_1} = \Xl{[-1:]}$ and $\Xl{}^{i_2,j_2} = \Xl{}$ which produces the column run constraint $\Xl {[-1:]}=\Xl{}$, \ie
$X_{i,j} = X_{i+1,j}$ for all corresponding values of $i$ and $j$. Once more, due to the existence of $\widetilde{\bfX}$ the same constraints holds due to $\Xr{}^{i_1,j_1} = \Xr{[-1:-1]}$ and $\Xr{}^{i_2,j_2} = \Xr{[:-1]}$.

In the center sub array $\Xc{}$ we need to distinguish whether $i_1 \leq i_2$ or $i_2 > i_1$, since this imposes different constraints on $\Xc{}$. For the first case we notice that $\Xc{}^{i_1,j_1} = \Xc{[-1:-1]}$ and $\Xc{}^{i_2,j_2}= \Xc{}$ which leads to a diagonal run constraint of $\Xc{[-1:-1]} = \Xc{}$, \ie $X_{i,j} = X_{i+1,j+1}$ for all corresponding values of $i$ and $j$. In the case where $i_2>i_1$, we see that $\Xc{}^{i_1,j_1}=\Xc{[:-1]}$ and $\Xc{}{^{i_2,j_2}} = \Xc{[-1:]}$. Therefore we need the anti-diagonal run constraint, \ie $X_{i,j} = X_{i+1,j-1}$ for all corresponding values of $i$ and $j$.

The constraints imposed on the sub arrays are exactly the same as the structure imposed on the array $\bfX$ which concludes the first part of the proof.

The reverse statement follows by observing that an array $\bfX$ satisfying the structure described in the claim will result in $\widetilde{\bfX} = \bfX^{i_1,j_1} = \bfX^{i_2,j_2}$. The reason is that this structure makes the sub arrays invariant to the different shifts in $\bfX$ resulting from both $(i_1,j_1)$ and $(i_2,j_2)$ \crisscross deletions.
\end{IEEEproof}

We use the following nomenclature throughout this section. 

\noindent\emph{Good and bad arrays:} An array $\bfX \in \sigmatnq$ is called \emph{good} \reply{if its deletion ball is larger than $\frac{2n^2}{5}$, i.e.,} 
\ExtVe{$|\dball_1(\bfX)| \geq \frac{2n^2}{5}$} and $\bfX$ is called \emph{bad} otherwise. Denote by $\cG_n,\cB_n$ the set of all good and bad arrays in $\sigmatnq$, respectively.

\noindent \reply{\emph{Bad columns and rows:} A column $\bfX_{[n],j}$, $j\in [2,n]$, is called \emph{bad} if for any pair of row indices $i_1,i_2 \in [n]$ with $i_1<i_2$ the columns $\bfX_{[n],j}$ and $\bfX_{[n],j-1}$ satisfy the following constraints:
\begin{enumerate}
   \item They are identical in the intervals $[1, i_1-1]$ and $[i_2+1, n]$, \ie $\bfX_{[i_1-1],j} = \bfX_{[i_1-1],j - 1}$ and $\bfX_{[i_2+1,n],j} = \bfX_{[i_2+1,n],j - 1}$.
    \item The column $\bfX_{[n],j}$ is either identical to $\bfX_{[n],j-1}$ up to a single down shift in the interval $[i_1,i_2]$, \ie $\bfX_{i+1,j} = \bfX_{i,j - 1}$ for all $i\in [i_1,i_2-1]$; or identical to $\bfX_{[n],j-1}$ up to a single up shift in the interval $[i_1,i_2]$, \ie $\bfX_{i-1,j} = \bfX_{i,j - 1}$ for all $i\in[i_1+1,i_2]$.
\end{enumerate}
For the case $i_1 = i_2$ a column $\bfX_{[n],j}$ is bad if it is identical to the column $\bfX_{[n],j - 1}$, except for the bit $i_1$ which can have an arbitrary value, i.e., $\bfX_{[i_1-1],j} = \bfX_{[i_1-1],j-1}$ and $\bfX_{[i_1+1,n],j} = \bfX_{[i_1+1,n],j-1}$. 
Columns that do not satisfy the aforementioned constraints are referred to as \emph{good} columns. Bad rows and good rows are defined similarly.}
%

\begin{claim}\label{claim:necessarily-good-arrays}
The deletion ball size of an array $\bfX \in \sigmatnq$ is bounded from below by $$|\dball(\bfX)| \geq \lvert \cI_\bfX^{c} \rvert \lvert \cI_\bfX^{r} \rvert.$$
Thus, $\bfX$ is good if the numbers of good columns and the number of good rows is at least $\sqrt{\frac{2}{5}}n$, i.e., $\lvert \cI_\bfX^{c} \rvert \geq \sqrt{\frac{2}{5}}n$ and  $\lvert \cI_\bfX^{r} \rvert \geq \sqrt{\frac{2}{5}}n$.
\end{claim}
\begin{IEEEproof}
Let $c_g = \lvert \cI_\bfX ^{c} \rvert$ and $r_g = \lvert \cI_\bfX ^{r} \rvert$ be the number of good columns and good rows, respectively. For a column $\bfX_{[n],j},\ j \in [n],$ the number of distinct arrays resulting from deleting the $j$-th column and any row $\bfX_{i,[n]}, i\in [n]$ is greater than or equal to $r_g$. In other words, deleting column $\bfX_{[n],j}$ and any good row $\bfX_{i_g,[n]},\ i_g \in \cI_\bfX ^{r}$ gives a new distinct array. To see this, assume by contradiction that there exists a pair $i_1,i_2\in \cI_\bfX^{r}$ such that $\bfX^{i_1,j} = \bfX^{i_2,j}$. Then, according to Lemma~\ref{lem:del-pattern}, all rows $\bfX_{i_1,[n]}$ up to $\bfX_{i_2,[n]}$ must be identical. Thus we have a contradiction since $\bfX_{i_2,[n]} \neq \bfX_{i_2 -1,[n]}$ by the definition of a good row.

We now turn our attention to good columns. For any $j_g \in \cI_\bfX^{c}$, the arrays resulting from deleting column $\bfX_{[n],j_g}$ and any row are distinct. Assume by contradiction that there exist $j_2 \in \cI_\bfX^{c}, j_2 \neq j_g,$ and $i_1,i_2 \in [n]$ such that $\bfX^{i_1,j_g} = \bfX^{i_2,j_2}$. Let $j_m = \max\{j_g,j_2\}$. Then according to Lemma~\ref{lem:del-pattern}, the columns $\bfX_{[n],j_m-1}$ and $\bfX_{[n],j_m}$ must satisfy $X_{i,j_m-1} = X_{i,j_m}$ for all $i\in [1,\min\{i_1,i_2\}-1] \cup [\max\{i_1,i_2\}+1,n]$ and $\bfX_{[n],j_2}$ must be identical to $\bfX_{[n],j_2}$ up to a single up or down shift in the interval $[\min\{i_1,i_2\},\max\{i_1,i_2\}]$ depending whether $i_1>i_2$ or $i_2>i_1$. However, this contradicts the definition of a good column. 

Hence, $|\dball (\bfX)| \geq c_g r_g$. Thus, if $c_g\geq \sqrt{\frac{2}{5}}n$ and $r_g\geq \sqrt{\frac{2}{5}}n$, then  $|\dball (\bfX)| \geq 2n^2/5$ and $\bfX$ is a good array.

\end{IEEEproof}
\begin{claim}\label{claim:numforbidden}
For $n \geq 5$ the number of possible choices of a good column (or row) \reply{is equal to \mbox{$\left(2^n - 2n^2\right)$}}.
\end{claim}
\begin{IEEEproof}
\reply{For a column $\bfX_{[n],j_b}$ to be bad, for any $i\in [n]$ it could be identical to column $\bfX_{[n],j_b-1}$ on all entries except the $i\th$ entry which can still be arbitrary, i.e., for all $i\in [n]$ it must hold that $\bfX_{[i-1],j_b}= \bfX_{[i-1],j_b-1}$ and $\bfX_{[i+1,n],j_b}= \bfX_{[i+1,n],j_b-1}$. There are $2n$ such columns. In addition, for two integers $i_1,i_2\in [n]$ such that $i_1<i_2$, a bad column $\bfX_{[n],j_b}$ could also be identical to $\bfX_{[n],j_b-1}$ on the intervals $[1,i_i-1]$ and $[i_2+1,n]$, i.e., $\bfX_{[i_1-1],j_b}=\bfX_{[i_1-1],j_b}$ and $\bfX_{[i_2+1,n],j_b}=\bfX_{[i_2+1],j_b}$; and identical to column $\bfX_{[n],j_b-1}$ up to a single down shift on the interval $[i_1,i_2]$, i.e., $\bfX_{[i+1,j_b} = \bfX_{i,j_b-1}$ for all $i\in[i_1,i_2-1]$. We have $2\binom{n}{2}$ such columns. Similarly there are $2\binom{n}{2}$ bad columns resulting from being identical to $\bfX_{[n],j_b-1}$ up to a single up shift on the interval $[i_1,i_2]$. Therefore, the total number of bad columns is $b_n\triangleq 2n+4\binom{n}{2} = 2n^2$ and the total number of good columns is equal to $2^{n^2}-2n^2$. The same calculation holds for good and bad rows.}
\end{IEEEproof}
We are ready to give an upper bound on $|\cB_n|$, the number of bad arrays.
\begin{lem}\label{lem:nbbad}
For $n\geq 41$ and $q \geq 2$ the number of bad arrays is bounded from above by 
\begin{align*}
\lvert \cB_n \rvert \leq \sqrt{\frac{8}{5}} \cdot q^{n^2-3n}.
\end{align*}
\end{lem}
\begin{IEEEproof}
If an array $\bfX \in \sigmatnq$ satisfies the conditions of Claim~\ref{claim:necessarily-good-arrays}, then it is a good array. Otherwise, $\bfX$ can be either a good array or a bad array. Therefore, we can compute an upper bound on the number of bad arrays as the number of arrays that do not satisfy the conditions of Claim~\ref{claim:necessarily-good-arrays}, i.e., have either less than $\sqrt{\frac{2}{5}}n$ good columns or less than $\sqrt{\frac{2}{5}}n$ good rows. Thus, we can write
\begin{align*}
    \lvert \cB_n \rvert &\leq 2 \sum _{j=n-\sqrt{\frac{2}{5}}n+1}^n \binom{n}{j} (b_n) ^j q^{n \cdot (n-j)} \\
    & \leq 2 \sqrt{\frac{2}{5}}n 2^n  (2 n^2 )^n q^{\sqrt{\frac{2}{5}}n^2 - n} = \sqrt{\frac{8}{5}}n  (4 n^2 )^n q^{\sqrt{\frac{2}{5}}n^2 - n} \\
    & = \sqrt{\frac{8}{5}} q^{\sqrt{\frac{2}{5}}n^2 - n+\log_q(n) +n\log_q(4n^2)}\\
    & \leq \sqrt{\frac{8}{5}} q^{\sqrt{\frac{2}{5}}n^2 - n+\log_2(n) +n\log_2(4n^2)}\\
    & \leq \sqrt{\frac{8}{5}} \cdot q^{n^2-3n},
\end{align*}
where \reply{\mbox{$b_n = 2n^2$}} results from the observation of Claim~\ref{claim:numforbidden}. 
\reply{The upper bound can be interpreted as summing over all arrays with at least $n-\sqrt{\frac{2}{5}}n+1$ bad columns (or bad rows) which can be located at $\binom{n}{j}$ different positions. The other columns (rows) can be chosen arbitrarily. The second inequality is obtained by bounding $\tbinom{n}{j}$ with $2^n$ and the last inequality holds for $n \geq 41$.}
\end{IEEEproof}

We now use the upper bound on the number of bad arrays to prove the following lower bound on the redundancy of a criss-cross deletion correcting code.

\begin{theorem}\label{thm:non-asymp-upper_bound}
The cardinality of any $q$-ary \crisscross deletion correcting code $\cC$ for $n\geq 41$ and $q\geq 2$ is bounded by
\begin{flalign*}
\lvert \cC \rvert \leq (1+\varepsilon) \frac{q^{n^2}}{q^{2n-1}\cdot \frac{2n^2}{5}},
\end{flalign*}
with \reply{$\varepsilon = 0.29$}, and thus its redundancy is lower bounded by
\begin{align*}
R \geq 2n-3+2\log_q (n) .
\end{align*}
\end{theorem}
\begin{IEEEproof}
Let $\cC_\cB \triangleq \cC\cap \cB_n$ and $\cC_\cG\triangleq \cC\cap \cG_n$. \ExtVe{Consider the following sphere packing argument }
\begin{align*}
\frac{2n^2}{5}\lvert \cC_\cG \rvert \leq \sum _{\bfX \in \cC_\cG } \lvert \dball _1 (\bfX) \rvert \leq \sum _{\bfX \in \cC } \lvert \dball _1 (\bfX) \rvert \leq q^{(n-1)^2}.
\end{align*}
Hence, $\lvert \cC_\cG \rvert \leq \frac{q^{(n-1)^2}}{\frac{2n^2}{5}}$. From Lemma~\ref{lem:nbbad}, for $n\geq \reply{41}$ the number of bad arrays is bounded by $\lvert \cB_n \rvert \leq \sqrt{\frac{8}{5}} \cdot q^{n^2-3n}$. Thus,
\begin{align}
    \lvert \cC \rvert & = \lvert \cC_\cG \rvert +\lvert \cC_\cB \rvert \nonumber \\
    & \leq \lvert \cC_\cG \rvert +\lvert \cB_n \rvert \nonumber \\
    & \leq \frac{q^{(n-1)^2}}{\frac{2n^2}{5}} + \sqrt{\frac{8}{5}} \cdot q^{n^2-3n} \nonumber \\
    & = \frac{q^{n^2}}{q^{2n-1}\cdot \frac{2n^2}{5}}\left(1 +\sqrt{\frac{32}{125}} \frac{n^2}{q^{n+1}} \right). \label{eq:card-expr}
\end{align}%
We show next that for any $n \geq 5$ and $q=2$ we have
\begin{align*}
    f(n) \triangleq \sqrt{\frac{32}{125}} \frac{n^2}{q^{n+1}} < \varepsilon = 0.29.
\end{align*}
Then we plug this result in~\eqref{eq:card-expr} to obtain the bound on the cardinality of $\cC$.

First, the only root of $f(n)$ is at $n=0$. Second, we can find the only maximum at $f(\frac{2}{\log_e(2)}) = 0.2850$ and a saddle point at $f(0) = 0$. Therefore, for $n \geq 3$ the function is monotonically decreasing and is always greater than zero. Moreover, it holds for any $n \geq 3$ and $q>2$
\begin{align*}
     \sqrt{\frac{32}{125}} \frac{n^2}{q^{n+1}} <  \sqrt{\frac{32}{125}} \frac{n^2}{2^{n+1}}  \leq f\left(\frac{2}{\log_e(2)}\right) < \varepsilon = 0.29,
\end{align*}
since $2^{n+1} < q^{n+1}$.

Consequently, we can re-write the upper bound in \eqref{eq:card-expr} with $n \geq 41$ and $\varepsilon = 0.29$ as
\begin{align*}
\lvert \cC \rvert &\leq (1+\varepsilon) \frac{q^{n^2}}{q^{2n-1}\cdot \frac{2n^2}{5}},
\end{align*}
which coincides with the expression in the theorem.

For the lower bound on the redundancy of a criss-cross code, we calculate a bound on $R = n^2 - \log_q (\lvert \cC \rvert)$ as
\begin{align*}
R &\geq n^2 - \log _q \left( (1+\varepsilon) \frac{q^{n^2}}{q^{2n-1}\cdot \frac{2n^2}{5}} \right) \\
&= n^2 -\log_q (1+\varepsilon) \\
& ~~~ - n^2  + (2n-1) + 2 \log _q (n) - \log_q\left(\frac{5}{2}\right) \\
&\geq 2n -3 + 2 \log_q (n),
\end{align*}
since $\log_q(1+\varepsilon) + \log_q (5/2) < \log_2(1.29) + \log_2(5/2) < 2$.
\end{IEEEproof}

In the following we write $f(n) \approx g(n)$ or $f(n) \lesssim g(n)$ if the equality or inequality holds when $n$ goes to infinity. 
\begin{cor}\label{cor:upper_bound}
For a \crisscross deletion correcting code $\cC$ and $n \to \infty$ the following holds:
\begin{flalign*}
\lvert \cC \rvert \lesssim \frac{q^{n^2}}{q^{2n-1}\cdot \frac{n^2}{2}},
\end{flalign*}
thus its asymptotic redundancy is at least $2n-2+2\log_q n$.
\end{cor}
\begin{IEEEproof}
The derivations are similar to the ones in Theorem~\ref{thm:non-asymp-upper_bound}. We only change the definition of good arrays to have deletion balls greater than or equal to $n^2/2$. A complete proof is given in Appendix~\ref{app:cor_asymptotic}.
%
\end{IEEEproof}

%% file: bad_matrix.tikz
\begin{tikzpicture}


\definecolor{color0}{rgb}{0.12156862745098,0.466666666666667,0.705882352941177}
\definecolor{color1}{rgb}{1,0.498039215686275,0.0549019607843137}
\definecolor{color2}{rgb}{0.172549019607843,0.627450980392157,0.172549019607843}
\definecolor{color3}{rgb}{0.83921568627451,0.152941176470588,0.156862745098039}
\definecolor{color4}{rgb}{0.580392156862745,0.403921568627451,0.741176470588235}
\definecolor{color6}{rgb}{0.549019607843137,0.337254901960784,0.294117647058824}
\definecolor{color5}{rgb}{0.890196078431372,0.466666666666667,0.76078431372549}

\definecolor{lightseagreen}{rgb}{0.13, 0.7, 0.67}
\definecolor{darkolivegreen}{rgb}{0.33, 0.42, 0.18}
\definecolor{mediumpersianblue}{rgb}{0.0, 0.4, 0.65}
\definecolor{bostonuniversityred}{rgb}{0.8, 0.0, 0.0}


\draw (0,0) rectangle (6,6);
\draw[pattern color=mediumpersianblue,pattern=vertical lines] (0,2) rectangle (2,4);
\node[circle, fill = white, inner sep = 0pt] at (1,3) {\textcolor{mediumpersianblue}{$\XL$}};

\draw[pattern color=color1,pattern=vertical lines] (4,2) rectangle (6,4);
\node[circle, fill = white, inner sep = 0pt] at (5,3) {\textcolor{color1}{$\XR$}};

\draw[pattern color=color4,pattern=horizontal lines] (2,0) rectangle (4,2);
\node[circle, fill = white, inner sep = 0pt] at (3,1) {\textcolor{color4}{$\XB$}};

\draw[pattern color=color6,pattern=horizontal lines] (2,4) rectangle (4,6);
\node[circle, fill = white, inner sep = 0pt] at (3,5) {\textcolor{color6}{$\XT$}};

\draw[pattern color=lightseagreen,pattern=north west lines] (2,2) rectangle (4,4);
\node[circle, fill = white, inner sep = 0pt] at (3,3) {\textcolor{black}{$\XC$}};

\node[circle, fill = white, inner sep = 0pt] at (0.4,0.4) {\Xiii{}};
\node[circle, fill = white, inner sep = 0pt] at (0.4,5.6) {\XI{}};
\node[circle, fill = white, inner sep = 0pt] at (5.6,0.4) {\Xiv{}};
\node[circle, fill = white, inner sep = 0pt] at (5.6,5.6) {\Xii{}};

\draw [->, color=gray] (2,6.5) -- (2,6.2) node [black,midway,yshift=0.4cm, color = gray] {$j_1$};
\draw [->, color=gray] (4,6.5) -- (4,6.2) node [black,midway,yshift=0.4cm, color = gray] {$j_2$};
\draw [->, color=gray] (-0.5,4) -- (-0.2,4) node [black,midway,xshift=-0.4cm, color = gray] {$i_1$};
\draw [->, color=gray] (-0.5,2) -- (-0.2,2) node [black,midway,xshift=-0.4cm, color = gray] {$i_2$};

\draw[ultra thick, color = bostonuniversityred] (2,6.2) -- (2,-0.2);
\draw[ultra thick, color = bostonuniversityred] (-0.2,4) -- (6.2,4);

\draw[ultra thick, color = darkolivegreen] (4,6.2) -- (4,-0.2);
\draw[ultra thick, color = darkolivegreen] (-0.2,2) -- (6.2,2);

\node[circle, fill = white, inner sep = 0pt] at (2,2) {\textcolor{color3}{?}};
\node[circle, fill = white, inner sep = 0pt] at (4,4) {\textcolor{color3}{?}};

\end{tikzpicture}

%% file: cons.tex
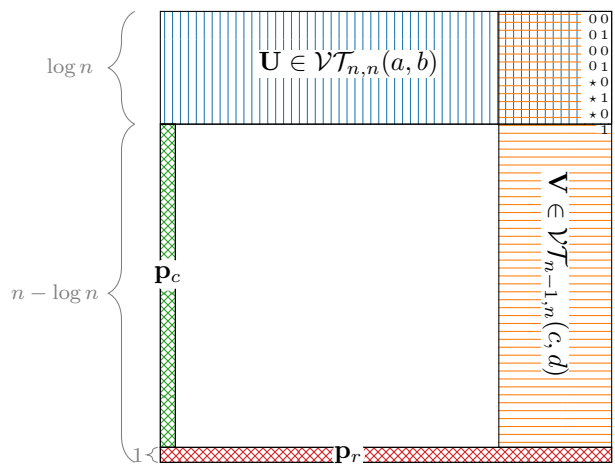
\begin{figure}[b!]
\centering
\input{Crisscross.tex}
\caption{The structure of the codewords of our \codename code. $\bfU$ is the binary representation of a $q$-ary vector $\mathbf{u} \in \VTab{q}$ with $q=n$. Each column is viewed as a symbol of the VT coded vector $\mathbf{u}$. The last column of $\bfU$ is an alternating sequence and the second to last column must start with four consecutive $0$'s. $\bfV$ is defined similarly to $\bfU$ where each row is a symbol of a VT coded vector $\mathbf{v}\in \VTcd{n}$. The alternating sequence of $\bfU$ is extended by one bit in $\bfV$. \ExtVe{For the encoding of $\bfV$, we replace the alternating sequence by the all $0$ sequence.} \reply{The column $\mathbf{p}_c$ is a parity column consisting of the sum of all columns of its size (and position). The row $\mathbf{p}_r$ is a parity row defined similarly to $\mathbf{p}_c$.} We denote by $\bfX \in\VTab{q}$ the binary representation of a $q$-ary vector $\mathbf{x} \in \Sigma_q^n$, such that $\mathbf{x}\in \VTab{q}$.}
\label{fig:encoding}
\end{figure}

In this section we present our \emph{\codename codes} that can correct a \crisscross deletion \ExtVe{or insertion} and state their main properties. Throughout the rest of the paper, we only consider binary arrays \reply{for the ease of presentation. The same construction can be extended for $q$-ary arrays.} We denote the set of all binary arrays $\{0,1\}^{n\times n}$ by $\sigmatn$. Moreover, we assume that $a,b,c,d$ are non-negative integers such that $0\leq a,b,d\leq n-1$ and $0\leq c\leq n-2$. We also assume that $n$ is a power of $2$ so that $\log n$ is an integer, while the extension for other values of $n$ will be clear from the context. The main results of this section are summarized in the following theorem and corollary.
\begin{theorem}\label{thm:cons}
The \codename code $\cC_{n}(a,b,c,d)$ (defined in Construction~\ref{const:crisscross}) is a \crisscross deletion \ExtVe{and insertion} correcting code that has an explicit decoder.
\end{theorem}

\begin{cor}\label{corr:red}
There exist integers $a,b,c,d$ for  which the  redundancy  of the \codename code $\cC_{n}(a,b,c,d)$ is at most $$2n+ \log n + 7 +2 \log e$$ bits and is therefore at most $ 2\log n+10+2\log e $ bits away from the lower bound.
\end{cor}

We prove Theorem~\ref{thm:cons} through a detailed explanation of the code construction that will be given in Section~\ref{subsec:cons}. In Section~\ref{subsec:dec}, we show how the decoding works \ExtVe{for a \crisscross deletion and insertion}. Afterwards, we compute an upper bound on the redundancy in Section~\ref{subsec:red}, and thus prove Corollary~\ref{corr:red}.

\subsection{The Construction} \label{subsec:cons}
The \codename code $\cC$ is an existential code whose codewords are $n\times n$ binary arrays structured as shown in Figure~\ref{fig:encoding} and as explained next. 
The code consists of two main components. The columns and the rows are indexed using the binary expansion $\bfU$ and $\bfV$ of two $n$-ary VT coded vectors to recover the positions of the \ExtVe{inserted/}deleted row and column. The parity bits $\mathbf{p}_c$ and $\mathbf{p}_r$ are used to recover the deleted information \ExtVe{in case of a deletion and to help detect the location of an inserted column or row in case of an insertion}.


{\em Indexing the columns: }The first $\log n$ rows of a codeword $\bfC\in \cC$ are the binary representation of a $q$-ary vector $\mathbf{u}$ encoded using a VT code $\VTab{q}$ that can correct one \ExtVe{insertion/}deletion, where $q=n$. The $\log n\times n$ binary array $\bfU$ satisfies the following requirements: \begin{enumerate*}[label=\textit{\roman*)}] \item every column of $\bfU$ is the binary representation of a symbol of the VT coded vector $\mathbf{u}\in \VTab{n}$; \item any two consecutive columns are different; \item the last column is the alternating sequence \ExtVe{that starts with $0$}; and \item the first $4$ bits of the second to last column are $0$'s.\end{enumerate*} As we shall see in the decoding section, this array serves as an index of the columns. That is, it allows the decoder to exactly recover the position of the \ExtVe{inserted/}deleted column.

{\em Indexing the rows: }The $(n-1)\times \log n$ array formed of the last $\log n$ bits of rows $1$ to $n-1$ (situated at the right of the array $\bfC$) is the binary representation of a $q$-ary vector $\mathbf{v}$ encoded using a $VT$ code $\VTcd{q}$ that can correct one \ExtVe{insertion/}deletion, with $q=n$. The $(n -1)\times \log n$ binary array $\bfV$ satisfies the following requirements: \begin{enumerate*}[label=\textit{\roman*)}] \item each row of $\bfV$ is the binary representation of a symbol of the VT coded vector $\mathbf{v}\in \VTcd{n}$; \item any two consecutive rows are different; \item the first $\log n$ rows also satisfy the requirements imposed on $\bfU$, with the exception of replacing the alternating sequence by the all $0$ sequence\footnote{The array $\bfU$ can still store the alternating sequence. When checking the constraints on the rows of $\bfV$ we assume the last column of $\bfU$ is the all $0$ sequence. Similarly, when encoding $\bfV$ we also assume that the last column is all $0$. Since this information is known by the decoder, the all $0$ sequence need not be stored in the array.
}; and \item the first bit below the alternating sequence is the opposite of the last bit of the alternating sequence. In other words, the alternating sequence is of length $\log n +1$. Again here we assume that the stored bit belongs to the alternating sequence, but for the encoding of $\mathbf{v}$ we assume that the first $\log n +1$ bits of the last column are all $0$'s.\end{enumerate*}  This array serves as an index of the rows that allows the decoder to recover the position of the \ExtVe{inserted/}deleted row.

{\em Parities: }The part of the first column of $\bfC$ that is not included in $\bfU$ is a parity of the same part of all corresponding columns, \ie each entry of that column is the sum of all bits corresponding to its same row. This column is denoted by $\mathbf{p}_c$ and is shown on the left in Figure~\ref{fig:encoding}. Moreover, the last row of $\bfC$ is a parity of all the rows and is denoted by $\mathbf{p}_r$. \ExtVe{In case of deletion, the parities allow the decoder to recover the information in the deleted column and row. In case of an insertion, the parities help the decoder to exactly recover the index of the inserted row and column in case the arrays $\bfU$ and $\bfV$ failed to do so, as explained in more details in the next section.} \reply{We start with an example to illustrate the idea before going into the formal definition of the construction. The example also illustrates the deletion decoder.}

\begin{example}
\reply{We construct a $9\times 9$ codeword of our code to illustrate the construction and the decoding algorithm. We chose $n$ to be $9$ (not a multiple of $2$) for convenience and to make the example simpler.

Assume that we want the columns and the rows to be indexed by codewords of $q$-ary VT codes with $q = 2^{(\left \lceil \log n \right \rceil)} = 16$ with $a= 2$, $b=0$, $c = 7$ and $d =0$. The first $4$ rows of the codeword should then be the binary representation of a $q$-ary vector $\mathbf{u}\in \mathcal{VT}_{9,16}(2,0)$. For clarity of presentation, we represent a symbol $\mathbf{x}=(x_1,x_2,x_3,x_4)^T \in \Sigma_{2^4}$ as the decimal representation $x = \sum_{i=1}^4x_i2^{i-1}$. Our construction requires the last symbol of $\mathbf{u}$ to be $10$, i.e., its binary representation is the alternating sequence. In addition, the second to last symbol of $\mathbf{u}$ must be $0$. Moreover, every two consecutive symbols of $\mathbf{u}$ must be different. An example is $\mathbf{u} = (0,1,2,3,4,5,11,0,10) \in \Sigma_{2^4}^9$. The binary representation $\bfU$ of $\mathbf{u}$ is the first four rows of $\bfX$ given in~\eqref{eq:codeword_ex}. Given $\mathbf{u}$, we now index the columns with a vector $\mathbf{v}\in \mathcal{VT}_{8,16}(7,0)$ such that the first four symbols of $\mathbf{v}$ are predetermined to be $3,10,1$ and $10$, respectively. Our construction requires the last bit of the binary representation of the fifth symbol of $\mathbf{v}$ to be set to $0$ as an extension of the alternating sequence of the last symbol of $\mathbf{u}$ (c.f.~\eqref{eq:codeword_ex}). Similarly to $\mathbf{u}$, any two consecutive symbols of $\mathbf{v}$ must be different. An example is $\mathbf{v} = (3,10,1,10,6,8,9,7) \in \Sigma_{2^4}^{8}$. The binary representation $\bfV$ of $\mathbf{v}$ is the transpose of the last four columns and first eight rows of the array $\bfX$ given in~\eqref{eq:codeword_ex}. The remaining entries of the array $\bfX$ not belonging the first column nor the last row (marked in black) are arbitrary. The entries of the last row (marked in red) are the column-wise parity bits. The remaining entries of the first column (in green) are the row-wise parity bits. The constructed codeword $\bfX$ is shown in~\eqref{eq:codeword_ex}.}
\definecolor{color0}{rgb}{0.12156862745098,0.466666666666667,0.705882352941177}
\definecolor{color1}{rgb}{1,0.498039215686275,0.0549019607843137}
\definecolor{color2}{rgb}{0.172549019607843,0.627450980392157,0.172549019607843}
\definecolor{color3}{rgb}{0.83921568627451,0.152941176470588,0.156862745098039}
\begin{equation}\label{eq:codeword_ex}
    \bfX = \begin{bmatrix}
        \color{color0} 0 & \color{color0} 1 & \color{color0} 0 & \color{color0} 1 & \color{color0} 0 & \color{color0} 1 & \color{color0} 1 & 0 &  0 \\
         \color{color0} 0 & \color{color0} 0 & \color{color0} 1 & \color{color0} 1 & \color{color0} 0 & \color{color0} 0 & \color{color0} 1 & 0 &  1 \\
          \color{color0} 0 & \color{color0} 0 & \color{color0} 0 & \color{color0} 0 & \color{color0} 1 & \color{color0} 1 & \color{color0} 0 & 0 & 0 \\
           \color{color0} 0 & \color{color0} 0 & \color{color0} 0 & \color{color0} 0 & \color{color0} 0 & \color{color0} 0 & \color{color0} 1 & 0 & 1 \\
            \color{color2} 0 & 0 & 0 & 0 & 0 & \color{color1} 0 & \color{color1} 0 & \color{color1} 1 & 0 \\
            \color{color2} 1 & 1 & 1 & 1 & 1 & \color{color1} 0 & \color{color1} 0 & \color{color1} 0 & \color{color1} 1 \\
            \color{color2} 0 & 0 & 1 & 0 & 1 & \color{color1} 1 & \color{color1} 0 & \color{color1} 0 & \color{color1} 1 \\
            \color{color2} 0 & 1 & 0 & 1 & 0 & \color{color1} 1 & \color{color1} 1 & \color{color1} 1 & \color{color1} 0 \\
            \color{color3} 1 & \color{color3} 1 & \color{color3} 1 & \color{color3} 0 & \color{color3} 1 & \color{color3} 0 & \color{color3} 0 & \color{color3} 0 & \color{color3} 0 \\
    \end{bmatrix},
     \qquad \qquad \bfX^{2,7} = \begin{bmatrix}
        \color{color0} 0 & \color{color0} 1 & \color{color0} 0 & \color{color0} 1 & \color{color0} 0 & \color{color0} 1 & \color{color0} 1 & 0 \\
          \color{color0} 0 & \color{color0} 0 & \color{color0} 0 & \color{color0} 0 & \color{color0} 1 & \color{color0} 1 & \color{color0} 0 & 0 \\
           \color{color0} 0 & \color{color0} 0 & \color{color0} 0 & \color{color0} 0 & \color{color0} 0 & \color{color0} 0 & \color{color0} 1 & 1 \\
            \color{color2} 0 & 0 & 0 & 0 & 0 & \color{color1} 0 & \color{color1} 0 & 0 \\
            \color{color2} 1 & 1 & 1 & 1 & 1 & \color{color1} 0 & \color{color1} 0 &  \color{color1} 1 \\
            \color{color2} 0 & 0 & 1 & 0 & 1 & \color{color1} 1 & \color{color1} 0 &  \color{color1} 1 \\
            \color{color2} 0 & 1 & 0 & 1 & 0 & \color{color1} 1 & \color{color1} 1 &  \color{color1} 1 \\
            \color{color3} 1 & \color{color3} 1 & \color{color3} 1 & \color{color3} 0 & \color{color3} 1 & \color{color3} 0 & \color{color3} 0 &  \color{color3} 0 \\
    \end{bmatrix}
\end{equation}

\reply{To illustrate the decoding strategy assume that column $7$ and row $2$ of $\bfX$ are deleted. The resulting array $\bfX^{2,7}$ is illustrated in~\eqref{eq:codeword_ex}. The decoder looks at the last non-deleted column and knows that it should be the alternating sequence because it is not the all zero column in the first four rows. From the last column, the decoder knows that the second row is deleted. Using the row-wise parity bits of the last row, the decoder can recover the second row. Now the decoder has all the rows of $\bfU$ with one deleted column and can thus use the VT decoder to recover the value and position of the lost column. Note that since every two consecutive columns in $\bfU$ are different, the decoder recovers the exact location of the deleted column. Having the index of the deleted column, the decoder recovers the values of the bits outside of $\bfU$ (rows $5$ to $8$) using the column-wise parity bits. The last bit of the deleted column is the sum of all other bits.}
\end{example}

%
%
%
Formally, the \codename code $\mathcal{C}$ can be seen as an intersection of four codes over $\sigmatn$ that define the constraints imposed on the codewords of $\mathcal{C}$. Let $\ell \triangleq \log n$, we define $\Alt$ to be the all zero array except for the first $\ell+1$ bits of the last column to be the alternating sequence, i.e., $\Alt_{[\ell+1],n} = [01010101 \cdots]^T$ and $W_{i,j} = 0$ otherwise. We denote by $\bfX \in\VTab{q}$ the binary representation of a $q$-ary vector $\mathbf{x} \in \Sigma_q^n$, such that $\mathbf{x}\in \VTab{q}$.

\begin{align*}
  \cU(a,b) &\triangleq \left\{\bfX:
  \begin{aligned}
  &\bfX_{[\ell],j} \neq \bfX_{[\ell],j+1} ,\quad j \in[n-1]\\
  &\bfX_{[4],n-1} = [0000]^T,\\
  &\bfX_{[\ell+1],n} = [010101\cdots]^T,\\
  &{\bfX_{[\ell],[n]}  \in \VTab{2^\ell}}
  \end{aligned}\right\},\\
 \hspace{-4ex} \cV(c,d) & \triangleq \left\{\bfX:
  \begin{aligned}
  &{\bfX}_{i,[n-\ell+1,n]} \neq {\bfX}_{i+1,[n-\ell+1,n]} , \ \ i\in[n-1] \\
  &\bfX_{[\ell+1],n} = [0000\cdots]^T,\\
  &\bfX_{[n-\ell+1,n],n-1}^T  \in \VTcd{2^\ell} 
  \end{aligned}\right\},\\
  \cV'(c,d) &\triangleq \left\{\bfY:
  \begin{aligned}
  &\bfY_{[\ell+1],n} = [010101\cdots]^T, \\
  &\bfY \oplus \Alt \in \cV(c,d)
  \end{aligned}\right\},\\
%
  \cP_c &\triangleq \left\{\bfX:
  x_{i,1} = \sum_{j=2}^n x_{i,j}, \quad  i=\ell+1,\dots,n-1
  \right\},\\
%
  \cP_r &\triangleq \left\{\bfX:
  x_{n,j} = \sum_{i=1}^{n-1} x_{i,j}, \quad  j \in [n] \right\} .
\end{align*}

\begin{construction}\label{const:crisscross}
The \codename code $\mathcal{C}_{n}(a,b,c,d)$ is the set of arrays $\bfC\in \sigmatn$ that belong to 
\begin{equation*}
    \mathcal{C}_{n}(a,b,c,d) \triangleq \cU(a,b) \cap \cV'(c,d) \cap \cP_c \cap \cP_r.
\end{equation*}
\end{construction}

\subsection{Decoder}\label{subsec:dec}
In this section, we show how the \codename code construction leverages the structure of a codeword $\bfC \in \mathcal{C}_{n}(a,b,c,d)$ to correct a criss-cross deletion \ExtVe{or insertion}. Formally, we prove Theorem~\ref{thm:cons}. \reply{We assume that the decoder knows the dimension of received array. In other words, the decoder knows whether a criss-cross deletion or a criss-cross insertion has happened and only needs to correct it.}

\noindent{\em Intuition: }The goal of the decoder is to use the \ExtVe{insertion/}deletion correction capability of $\VTab{n}$ and $\VTcd{n}$ to recover the positions of the \ExtVe{inserted/}deleted column and row. The decoder first uses the alternating sequence to check if a row of $\bfU$ is \ExtVe{inserted/}deleted and therefore \ExtVe{corrects} it before proceeding to the VT code decoder. \ExtVe{In case of a deletion,} the second to last column allows the decoder to detect whether the alternating sequence was deleted or not. If the alternating sequence is deleted, the decoder cannot use $\bfU$ and has to start using $\bfV$ to detect the position of the deleted row, recover it using the parities and then obtain the position of the deleted column from $\bfU$. The parities are used to recover the deleted information once the decoder has the position of the deleted row and/or column. \ExtVe{In case of an insertion, the inserted vector may be equal to another consecutive vector in either $\bfU$ or $\bfV$. In this case, the decoder uses the parity bits over the remaining part of the vector to exactly recover the position of the inserted row or column.} We are now ready to present the proof of Theorom~\ref{thm:cons}. 

\begin{IEEEproof}[Proof of Theorem~\ref{thm:cons}]
\ExtVe{We split the proof into two parts: \begin{enumerate*}[label={\emph{\alph*)}}] \item an explicit decoder for a \crisscross deletion; and \item an explicit decoder for a \crisscross insertion.
\end{enumerate*}}

\paragraph{Deletion correcting decoder}
The decoder for $ \mathcal{C}_{n}(a,b,c,d)$ receives as input an $(n-1) \times (n-1)$ array $\widetilde{\bfC}$ resulting from a \crisscross deletion in an array $\bfC$ of $ \mathcal{C}_{n}(a,b,c,d)$ and works as follows. The decoder starts by looking at the first $\ell \times (n-1)$ subarray of $\widetilde{\bfC}$ and examining the last column.

\indent {\em Case 1:} Assume the last column of $\bfC$ is not deleted. Using the alternating sequence, the decoder can detect whether or not there was a row deletion in $\bfU$ and locate its index. This is done by locating a run of length $2$ in the alternating sequence. The last bit of the alternating sequence falling in $\bfV$ and not in $\bfU$ ensures that the decoder can detect whether the last row of $\bfU$ is deleted or not.

\indent \indent {\em Case 1 (a): }If there was a row deletion in $\bfU$, the decoder uses the non deleted part of $\mathbf{p}_r$ to recover the deleted row \ExtVe{except for the bit in the deleted column}. The decoder can now use the properties of $\VTab{n}$ to decode the column deletion in $\bfU$. Since any two consecutive columns in $\bfU$ are different, the decoder can locate the exact position of the deleted column and recover its value. The position of the deleted column in $\bfU$ is the same as the deleted column in the whole array. Using $\mathbf{p}_c$, the decoder can now recover the remaining part of the deleted column.

\indent \indent {\em Case 1 (b): }If the deleted row was not in $\bfU$, the decoder uses $\VTab{n}$ to recover the index of the deleted column and its value within $\bfU$ and uses $\mathbf{p}_c$ to recover the value of the deleted column outside of $\bfU$ \ExtVe{(except for the bit in the intersection of the deleted row and column)}. Then, the decoder uses $\VTcd{n}$ to recover the index of the deleted row. Again, since any two consecutive rows in $\bfV$ are different the decoder can recover the exact position of the deleted row. Using $\mathbf{p}_r$, the decoder recovers the value of the bits of the deleted row.

\indent {\em Case 2: }Now assume that the last column of $\bfC$ is deleted. By looking at the last column of $\widetilde{\bfC}$, the decoder knows that the alternating sequence is missing thanks to the run of $0$'s inserted in the beginning of the second to last column of $\bfC$. Note that irrespective of the location of the row deletion, the last column will have a run of at least three $0$'s which cannot happen in the alternating sequence. Therefore, the decoder knows that the last column is deleted and starts by looking at $\bfV$. Using the parity $\mathbf{p}_c$, the decoder recovers the missing part of the deleted column that is in $\bfV$ but not in $\bfU$. By construction, the first $\ell$ bits of the last column of $\bfV$ are set to $0$ when encoding $\bfV$ using a VT code. 
Thus, the decoder recovers the whole missing column. By using the property of $\VTcd{n}$
, the decoder recovers the index of the missing row and uses $\mathbf{p}_r$ to recover the value of the  bits of this row. After recovering the deleted row the decoder adds the alternating sequence to $\bfU$ and recovers the whole array $\bfC$.

\paragraph{Insertion correcting decoder}
\ExtVe{The decoder for $ \mathcal{C}_{n}(a,b,c,d)$ receives as input an $(n+1) \times (n+1)$ array $\widetilde{\bfC}$ resulting from a \crisscross insertion of an array $\bfC$ of $ \mathcal{C}_{n}(a,b,c,d)$ and works as follows. The decoder starts by looking at the first $(\ell+1) \times (n+1)$ subarray of $\widetilde{\bfC}$ {(recall that $\ell=\log n$)} and examines the last two columns.}

\indent \ExtVe{{\em Case 1:} Assume the second to last column is not an alternating sequence (special insertions that we consider in cases 2 and 3) and the last column is the alternating sequence. Using the alternating sequence, the decoder can detect whether or not there was a row insertion in $\bfU$. This is done by locating a run of length $2$ in the alternating sequence.}

\indent \indent \ExtVe{{\em Case 1 (a): }If there was a row insertion in $\bfU$, the decoder has two candidates for the inserted row: the ones that cause the run of length $2$ in the alternating sequence. Recall that the bit-wise sum of the $n$ rows of $\bfC$ is known to the decoder (parity check constraint). The decoder verifies which of the two candidate rows does not satisfy the parity constraints, i.e., the decoder sums the $n-1$ remaining rows together with each of the candidate rows and checks the Hamming weight of the resulting vector. The row that results in a vector with Hamming weight more than $1$ is the inserted row\footnote{\reply{The original row can only result in at most one $1$ located in the position of the inserted column.}}.
If both resulting vectors are different and result in Hamming weight $1$, then the decoder is confused between two candidates for the inserted row and two candidates for the inserted column. In this case, the decoder deletes both candidate rows and both candidate columns where a ``1'' appears in the resulting vectors and uses the deletion correction capability of the code to recover the original message. {This works since the inserted row and column were removed, i.e., the array is now affected by \emph{one} row deletion and \emph{one} column deletion.}} 

\indent \indent \ExtVe{Otherwise, the decoder removes the inserted row and uses the properties of $\VTab{n}$ to decode the column insertion in $\bfU$. Since any two consecutive columns in $\bfU$ are different, the inserted column is either different from both adjacent columns or equal to only one of them. In the former case, the decoder recovers the exact position of the inserted column and removes it. The position of the inserted column in $\bfU$ is the same as the inserted column in the whole array. In the latter case, the decoder has two candidates of inserted columns. The decoder uses the column parity check to verify which column is the inserted one and removes it. Note that since the inserted row is removed, the decoder will have at most one column that does not satisfy the column parity check constraints. If both columns verify the parity constraints, then they are identical.}

%
%

\indent \indent \ExtVe{{\em Case 1 (b): }If the inserted row was not in $\bfU$, i.e., the alternating sequence is intact, the decoder uses $\VTab{n}$ to recover the index and value of the inserted column in $\bfU$. If this column in $\bfU$ is different from both of its adjacent columns in $\bfU$, then the decoder removes the whole column and proceeds to correcting the inserted row. However, if the inserted column in $\bfU$ is equal to one of its adjacent columns (since any two consecutive columns are different), then the decoder has two candidates of inserted columns. In a similar way to Case~1~(a), the decoder uses the column parity check constraints to verify which column is the inserted one. After removing the inserted column, the decoder uses $\VTcd{n}$ to recover the index of the inserted row. Again, if the inserted row in $\bfV$ is different from both adjacent rows in $\bfV$, the decoder removes the whole row. Otherwise, the decoder has two candidates for the inserted rows; therefore the decoder uses the row parity check to recover the exact position of the inserted row.}


\indent \ExtVe{{\em Case 2: }Now assume that the two last columns of $\bfU$ are identical. Due to the $4$ zeros in the second to last column of $\bfC$ (now third to last column in $\widetilde{\bfC}$), the decoder detects that a column insertion happened in one of the last two columns of $\bfU$. The decoder uses the column parity check to verify which column is the inserted one. In case both columns satisfy the parity check constraints, then they are identical. If both columns violate the parity check constraints in one position, similarly to Case~1~(a), the decoder deletes both columns and both rows where the columns do not satisfy the parity check constraint and uses the deletion correction capability of the code. After removing the inserted column, the decoder examines the alternating sequence to check if the inserted row is in $\bfU$. If this is the case, the decoder uses the row parity check to verify which row is inserted and removes it. If the inserted row is not in $\bfU$, the decoder uses $\VTcd{n}$ and the column parity check to recover the exact index of the inserted row.}

\indent \ExtVe{{\em Case 3: }Assume that the last column of $\bfU$ is not the alternating sequence. Thus, the last column is an inserted column. The decoder removes this column and proceeds to detect which row is inserted as explained in the previous case.}
\end{IEEEproof}

\subsection{Redundancy of the code} \label{subsec:red}
The redundancy $R_{\mathcal{C}_{n}(a,b,c,d)}$ of $\mathcal{C}_{n}(a,b,c,d)$ is given by 
\begin{align*}
    R_{\mathcal{C}_{n}(a,b,c,d)} &= \log(2^{n^2}) - \log|\mathcal{C}_{n}(a,b,c,d)|\\
    & = n^2 - \log \left| \cU(a,b) \cap \cV'(c,d) \cap \cP_c \cap \cP_r \right|.
\end{align*}
In this section we show that there exist $a,b,c,d$ for which $$R_{\mathcal{C}_{n}(a,b,c,d)} \leq 2n + {4\log n} + 7 + 2\log e.$$  We do so by computing a lower bound on \mbox{$\log \left|\mathcal{C}_{n}(a,b,c,d)\right|$}. To that end we count the number of $n \times n$ binary arrays that satisfy all the requirements imposed on the codewords $\bfC$ in $\mathcal{C}_{n}(a,b,c,d)$. 

Since the constraints imposed on the codes $\cU(a,b) \cap \cV'(c,d)$, $\cP_c$, and $\cP_r$ are disjoint, we have that
\begin{align}
    R_{\mathcal{C}_{n}(a,b,c,d)} &= R_{\cU(a,b) \cap \cV'(c,d)} + R_{\cP_c} + R_{\cP_r} \nonumber\\
     &= R_{\cU(a,b) \cap \cV'(c,d)} + 2n-\log n-1.\label{eq:partities}
\end{align}
Equation~\eqref{eq:partities} follows from the fact that the $n-\log n-1$ bits of $\mathbf{p}_c$ and the $n$ bits of $\mathbf{p}_r$ are fixed to predetermined values.

We now compute an upper bound on the redundancy of the set $\cU(a,b) \cap \cV'(c,d)$.

\begin{proposition}\label{prop:uv}
There exists four values $a^\star, b^\star, c^\star$ and $d^\star$ for which the redundancy $R_1$ of $\cU(a^\star,b^\star) \cap \cV'(c^\star,d^\star)$ is bounded from above by
\begin{align}\label{eq:sets}
R_1< (2n-2\log n-3) \log \left(\frac{n}{n-1}\right) + 5\log n + 6.
\end{align}
\end{proposition}
From Equations~\eqref{eq:partities} and~\eqref{eq:sets} we obtain,
%
%
\begin{align}
    R_{\mathcal{C}_{n}(a,b,c,d)} & < (2n-2\log n-3) \log \left(\frac{n}{n-1}\right)+ 2n + 4\log n + 5 \nonumber\\ 
    & < 2n \log \left(\frac{n}{n-1}\right) + 2n + 4\log n + 5 \nonumber \\
    &< 2n + 4\log n + 5 + 2\log 2e \label{eq:loge}\\
    & = 2n + 4\log n + 7 + 2\log e. \nonumber
\end{align}
In~\eqref{eq:loge} we use the inequality {$2n\log \left(\frac{n}{n-1}\right) \leq 2\log 2e.$} \reply{This inequality follows from noting that $\left(1-\frac{1}{n}\right)^n$ is an increasing function of $n$ that converges to $\frac{1}{e}$ and is always greater than or equal to $0.25>\frac{1}{2e}=0.1852$ for $n\geq 2$.} The proof of Corollary~\ref{corr:red} is now complete. We conclude this section with the proof of Proposition~\ref{prop:uv}.
\begin{IEEEproof}[Proof of Proposition~\ref{prop:uv}]
We start with counting the number of arrays that satisfy all the imposed constraints except for the VT constraints in the codes $\cU(a^\star,b^\star)$ and 
$\cV'(c^\star,d^\star)$. To that end, we define the following three sets over $\sigmatn$. 
\begin{align*}
  \cU_\bot &\triangleq \left\{\bfX:
  \bfX_{[\ell],j} \neq \bfX_{[\ell],j+1}, \quad  j\in [n-\ell-1] \right\},\\
  \cV_\bot & \triangleq \left\{\bfX:
  \begin{aligned}
  &\bfX_{i, [n-\ell+1,n]} \neq \bfX_{i+1, [n-\ell+1,n]} , \quad \ell<i<n-1 \\
  &{X_{\ell+1,n}} \equiv \ell \mod 2\\
  \end{aligned}\right\},\\
%
  \cS_\cap &\triangleq \left\{\bfX:
  \begin{aligned}
  &\bfX_{[\ell],j} \neq \bfX_{[\ell],j+1} , \quad n-\ell\leq j<n,\\
  &\bfX_{i,[n-\ell+1,n]} \neq \bfX_{i+1,[n-\ell+1,n]}, i\in[\ell] ,\\
  &\bfX_{[4],n-1} = [0000]^T,\\
  &\bfX_{[\ell],n} = [010101\cdots]^T\\
  \end{aligned}\right\}.
\end{align*}

$\cU_\bot$ is the set of all $n\times n$ arrays in which any two consecutive columns, from column $1$ to $n-\ell$, are different when restricted to the first $\ell$ entries; $\cV_\bot$ is the  the set of all $n\times n$ arrays in which the entry $X_{\ell +1,n}$ is fixed to a predetermined value and any two consecutive rows, from row $\ell+1$ to $n-1$, are different when restricted to the last $\ell$ entries; and $\cS_\cap$ is the set of $n\times n$ arrays in which the $\ell \times \ell$ sub array ending at the last bit of the first row of the original array has distinct consecutive columns, distinct consecutive rows, the last row fixed to a predetermined value and the first $4$ bits of the second to last column are also predetermined. $\cS_\cap$ is also defined to guarantee that the first column of the $\ell \times \ell$ sub array is different from the $\ell$ entries of column $n-\ell$ and similarly to the last row.

\begin{claim}\label{claim:rect}
The redundancies of $\cU_\bot$ and $\cV_\bot$ are respectively {given by}
\begin{align*}
    R_{\cU_\bot} & = (n-\log n -1) \log \left( \frac{n}{n-1}\right),\\
    R_{\cV_\bot} &  = (n-\log n -2) \log \left( \frac{n}{n-1}\right) + 1.
\end{align*}
\end{claim}
The intuition behind Claim~\ref{claim:rect} is that the first $\log n$ bits of any two consecutive columns of $\bfU$ (last $\log n$ bits of any two consecutive rows of $\bfV$) must be different. The proof of Claim~\ref{claim:rect} is given in Appendix~\ref{app:cons_claims}.

\begin{claim}\label{claim:square}
The redundancy of $\cS_\cap$ is upper bounded by
\begin{equation}
    R_{\cS_\cap} < \log n + 5.
\end{equation}
\end{claim}

The intuition behind Claim~\ref{claim:square} is that with at most one bit of redundancy we can guarantee that every two consecutive rows and every to consecutive columns of the $\log n \times \log n$ square are different. The remaining $\log n +4$ bits are due to the use of the alternating sequence and fixing four bits of the second to last column of the square. The proof of Claim~\ref{claim:square} is given in Appendix~\ref{app:cons_claims}.

The remaining part of the proof is to count the number of arrays that satisfy the above requirements and have $\bfU \in \VTab{n}$ and $\bfV \in \VTcd{n}$. Using the same arguments explained in Section~\ref{sec:def}, we note that the VT constraints partition the set $\cU_\bot \cap \cV_\bot \cap\cS_\cap$ into $(n^3)(n-1)$ disjoint cosets.
 Therefore, there exist $a^\star,d^\star,c^\star,d^\star$ for which
\begin{align*}
|\cU(a^\star,b^\star) \cap \cV(c^\star,d^\star)| \geq \dfrac{|\cU_\bot \cap \cV_\bot \cap\cS_\cap|}{(n^3)(n-1)}.
\end{align*}
In other words, the redundancy $R_1$ of $\cU(a^\star,b^\star) \cap \cV(c^\star,d^\star)$ is bounded from above by
\begin{align*}
R_1 \leq R_{\cU_\bot \cap \cV_\bot \cap\cS_\cap} + \log \left((n^3)(n-1)\right).
\end{align*}

Since all the constraints in $\cU_\bot$, $\cV_\bot$, $\cS_\cap$ are disjoint by construction, we can rewrite the previous equation as
\begin{align}
R_1 & \leq R_{\cU_\bot} +  R_{\cV_\bot} + R_{\cS_\cap} + \log \left(n^3(n-1)\right) \nonumber\\
& < R_{\cU_\bot} +  R_{\cV_\bot} + R_{\cS_\cap} + 4\log n \label{eq:log}\\
& \leq (2n-2\log n-3) \log \left(\frac{n}{n-1}\right) + 5\log n + 6. \label{eq:claims}
\end{align}


In~\eqref{eq:claims} we substituted the results from Claim~\ref{claim:rect} and Claim~\ref{claim:square}.
\end{IEEEproof}

%% file: Crisscross.tex
\begin{tikzpicture}


\definecolor{color0}{rgb}{0.12156862745098,0.466666666666667,0.705882352941177}
\definecolor{color1}{rgb}{1,0.498039215686275,0.0549019607843137}
\definecolor{color2}{rgb}{0.172549019607843,0.627450980392157,0.172549019607843}
\definecolor{color3}{rgb}{0.83921568627451,0.152941176470588,0.156862745098039}

\draw[pattern color=color0,pattern=vertical lines] (0,0) rectangle (6,-1.5);
\draw[pattern color=color1,pattern=horizontal lines] (4.5,0) rectangle (6,-5.8);
\draw[pattern color=color2,pattern=north west lines] (0,-1.5) rectangle (0.2,-5.8);
\draw[pattern color=color2,pattern=north east lines] (0,-1.5) rectangle (0.2,-5.8);
\draw[pattern color=color3,pattern=north west lines] (0,-6) rectangle (6, -5.8);
\draw[pattern color=color3,pattern=north east lines] (0,-6) rectangle (6, -5.8);

\node[rectangle, fill = white, inner sep = 0pt] at (2.5, -0.7) {$\mathbf{U}\in \VTab{n}$};
\node[rectangle, fill = white, inner sep = 0pt, rotate = 270] at (5.25, -3.5) {$\mathbf{V}\in \VTcd{n}$};;
\node[circle, fill = white, inner sep = 0pt] at (0.1,-3.5) {$\mathbf{p}_c$};
\node[circle, fill = white, inner sep = 0pt] at (2.5,-5.9) {$\mathbf{p}_r$};

\draw[fill = white, draw = none] (5.6,-1.48) rectangle (5.99,-0.01);
\draw[fill = white, draw = none] (5.8,-1.61) rectangle (5.99,-1.51);

\draw (0,-1.5) -- (6,-1.5);

\node[inner sep = 0pt, anchor = north east, font = \tiny] at (6.15,0) {$\begin{array}{c}0\\1\\0\\1\\0\\1\\ 0 \\ 1\end{array}$};
\node[inner sep = 0pt, anchor = north east, font = \tiny] at (6,0) {$\begin{array}{c}0\\0\\0\\0\\\star \\ \star \\ \star \end{array}$};

\draw [decorate,decoration={brace,amplitude=10pt,mirror,raise=4pt},yshift=0pt, color = gray]
(-0.2,0) -- (-0.2,-1.5) node [black,midway,xshift=-1cm, color = gray] {\footnotesize
$\log n$};

\draw [decorate,decoration={brace,amplitude=10pt,mirror,raise=4pt},yshift=0pt, color = gray]
(-0.2,-1.5) -- (-0.2,-6) node [black,midway,xshift=-1.2cm, color = gray] {\footnotesize
$n-\log n$};

\draw [decorate,decoration={brace,amplitude=4pt,raise=1pt, mirror},yshift=0pt, color = gray]
(0,-5.8) -- (0,-6) node [black,midway,xshift=-0.3cm, color = gray] {\footnotesize
$1$};

\end{tikzpicture}

%% file: systematic.tex
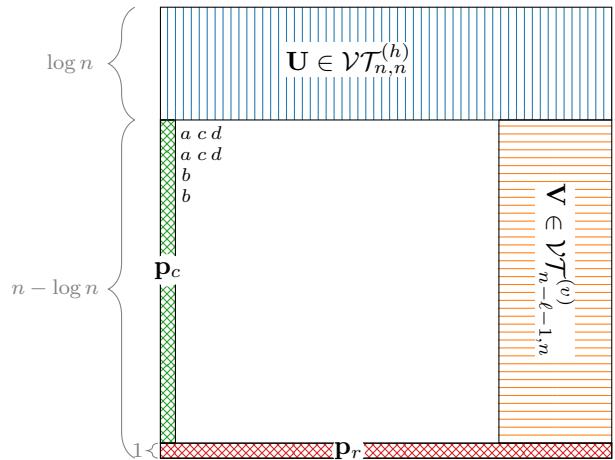
\begin{figure}[t!]
\centering
\input{Crisscross_explicit.tex}
\caption{The structure of the codewords of our \codename code with explicit encoder and decoder. $\bfU$ is the binary representation of a $q$-ary vector $\mathbf{u}$ encoded using an explicit VT code $\VTabs{q}$ with $q=n$. Each column is viewed as a symbol of the VT coded vector $\mathbf{u}$. $\bfV$ is defined similarly to $\bfU$ where each row is a symbol of a VT coded vector $\mathbf{v}$ encoded using an explicit VT code $\VTcds{n}$, where $\ell = \log n$. $\mathbf{p}_c$ is a parity column consisting of the sum of all columns of its size (and position). $\mathbf{p}_r$ is a parity row consisting of the sum of all rows. The first four bits of the second column below $\bfU$ (shown as $a,a,b,b$) are reserved to help the decoders of $\VTabs{n}$ and $\VTcds{n}$. The first two bits of the third and fourth columns below $\bfU$ (i.e., $c,c,d,d$) are reserved to help the decoder of $\VTabs{n}$. Choosing the values of $a$, $b$, $c$ and $d$ will be clarified.}
\label{fig:encoding_exp}
\end{figure}

In this section we show how to construct a \codename code with explicit encoder and decoder at the expense of increasing the redundancy by $5\log n+ 5$ bits. The main idea is to change the arrays $\bfU$ and $\bfV$ so that they are the binary representations of two $q$-ary vectors $\mathbf{u}$ and $\mathbf{v}$, which are encoded using two variations of the explicit systematic non-binary VT codes from~\cite{tenengolts1984nonbinary} and will be introduced in the sequel. The new structure of the codewords is depicted in Figure~\ref{fig:encoding_exp}. In the remaining of this section we also take $\log n$ to be an integer. The main result of this section is stated in the next theorem.
\begin{theorem}\label{thm:exp_cons}
The \codename code defined below, constructed by modifying Construction~\ref{const:crisscross}, is a \crisscross deletion and insertion correcting code that has explicit encoder and decoder. The redundancy of this code is bounded from above by
\begin{align*}
    R_{\text{explicit}} & < 2n + 9 \log n + 12 + 2\log e.
\end{align*}
\end{theorem}

\subsection{Construction}
We first review the non-binary systematic VT construction from~\cite{tenengolts1984nonbinary}.
\subsubsection{Systematic VT code} In \cite{tenengolts1984nonbinary} Tenengolts presented two VT code constructions: an existential construction as in Section~\ref{sec:def}, i.e., one defines the constraints on the codewords and shows that such a code exists; and a systematic construction that takes as input a message and only adds redundancy to it such that the resulting codeword satisfies some imposed constraints. Before going into the details of our construction, we explain the construction of the systematic VT code as presented in \cite{tenengolts1984nonbinary}.

The systematic VT code, denoted by $\VTs{q}$, takes as input a message $\mathbf{a} = (a_1,\dots, a_k) \in \Sigma_q^k$ and encodes it into a vector $\mathbf{c}\in \Sigma_q^n$ where\footnote{In the construction by Tenengolts, three 
extra symbols are added at the end of the sequence to account for the case of sending several concatenated codewords. We do not need those symbols here as only one array is sent through the channel.} $n= k + 3 + \lceil \log_q k \rceil$. Since in our case we take $q=n$, we explain here the construction of $\VTs{n}$, and $n=k+4$.
Given the message $\mathbf{a}$, the encoded vector $\mathbf{c}=(c_1,\dots,c_n) \in \Sigma_n^n$ of $\VTs{n}$ is constructed as follows. 
\begin{enumerate}
    \item The first $k$ symbols of $\mathbf{c}$, referred to as the {\em systematic data part}, are the same as the first $k$ symbols of $\mathbf{a}$, i.e., $c_i = a_i$ for $i=1,\dots,k$. 
    \item The symbols $c_{k+1}$ and $c_{k+2}$ satisfy $c_{k+1}= c_{k+2} = a_{k}+1\mod n$. 
    \item To compute $c_{k+3}$, the signature vector $\mathbf{s}=(s_1,\dots,s_k)$ is computed as $s_1 = 1$ and
    \begin{equation*}
      s_i = \begin{cases}
      1 & \hfill \text{ if } a_{i}\geq a_{i-1}\\ 
      0 & \hfill \text{ otherwise.}
      \end{cases}
    \end{equation*}
     The symbol\footnote{In the general $\VTs{q}$ where $q < n$, one needs $r= \lceil \log_q k \rceil$ symbols $c_{k+3},\dots,c_{k+3+r}$ to be the $q$-ary representation of the equal to $\sum_{i=1}^k (i-1)s_i \mod k$.} $c_{k+3}$ is then equal to $\sum_{i=1}^k (i-1)s_i \mod k$. 
    \item The symbol $c_{k+4}$ is computed as $c_{k+4} = \sum_{i=1}^k c_i \mod n$. 
\end{enumerate}

The symbols $c_{k+3}$ and $c_{k+4}$ are referred to as the {\em parity symbols}. 
The symbols $c_{k+1}$ and $c_{k+2}$ are used as separators between the data part and the parity part so that the decoder can localize the insertion/deletion. Note that they can have other values besides $a_{k}+1\mod n$ as long as they are different from $a_k$. We will use this variation in our construction. If the insertion/deletion happens in the data part, the decoder uses $c_{k+3}$ and $c_{k+4}$ together with the same VT decoder explained in \cite{tenengolts1984nonbinary} to decode the insertion/deletion. Otherwise, the data part is intact and no decoding is needed.

\subsubsection{Encoding of $\bfU$ and $\bfV$} We slightly modify the systematic VT code to fit our setting. Namely, for the vector $\mathbf{u}$ (used to compute the array $\bfU$) we put the systematic part of the data in the end of the sequence and the parity part in the beginning. For the vector $\mathbf{v}$, we maintain the structure of the systematic VT code. For both vectors $\mathbf{u}$ and $\mathbf{v}$ we pre-encode the message so that every two consecutive symbols of the vectors $\mathbf{u}$ and $\mathbf{v}$ are different. Thus, the construction will not be systematic, but explicit. Furthermore, we also change the separator symbols to better fit our setting. We require any two consecutive symbols to be different to detect a row deletion within $\bfU$. These modifications require also adding one more redundancy symbol.

\paragraph{Horizontal VT encoder $\VTabs{n}$} Consider the message \mbox{$\mathbf{a} = (a_1,\dots,a_k) \in \{1,\dots,n-1\}^k$} to be encoded into the vector $\mathbf{u}\in \Sigma_n^n$. Here we take $n=k+5$. \reply{For notational convenience we number the indices of $\mathbf{u}$ from $-4$ to $n-5$, i.e., $\mathbf{u} = (u_{-4},u_{-3},\dots, u_{n-5})$. }The vector $\mathbf{u}$ is constructed as follows.
\begin{enumerate}
    \item To guarantee that every two consecutive symbols are different, the symbol \reply{$u_1$} is made equal to $a_1$ and the symbols \reply{$u_2$} to \reply{$u_{n-5}$} are computed as \reply{$u_i = u_{i-1} + a_{i} \mod n$}. 
    \item The symbol \reply{$u_0$} can take an arbitrary value up to the restrictions explained in the sequel. 
    \item The symbol \reply{$u_{-4}$} is computed as \reply{$u_{-4} = \sum_{i=0}^{n-5} u_i \mod n$}. 
    \item To compute \reply{$u_{-3}$}, we compute the signature vector $\mathbf{s}=(s_1,\dots,s_{k+1})$ as $s_1 = 1$ and for $i=2,\dots,n$
    \begin{equation*}
      s_i = \begin{cases}
      1 & \hfill \text{ if }  \reply{u_{i-1}\geq u_{i-2}}\\ 
      0 & \hfill \text{ otherwise.}
      \end{cases}
    \end{equation*} The symbol \reply{$u_{-3}$} is then equal to $\sum_{i=1}^{k+1}(i-1)s_i \mod (k+1)$. 
    \item The symbol \reply{$u_{-2}$} is the $n$-ary value of a length $\log n$ alternating sequence and \reply{$u_{-1}$} is chosen as the complement alternating sequence of the binary representation of \reply{$u_{-2}$}. 
    \item Since for our \codename code we need any two consecutive symbols to be different, $u_i\neq u_{i \pm 1}$ for all $i$, we choose the value of \reply{$u_0$} to be different from \reply{$u_{-1}$} and \reply{$u_1$} such that it ensures that \reply{$u_{-4}\neq u_{-3}$} and \reply{$u_{-3}\neq u_{-2}$}. This is proved in the next claim.
\end{enumerate}


\begin{claim}\label{claim:u5}
A value for \reply{$u_0$} that satisfies \reply{$u_0\neq u_{-1}$}, \reply{$u_0\neq u_1$} and makes \reply{$u_{-4}\neq u_{-3}$} and \reply{$u_{-3}\neq u_{-2}$} always exists.
\end{claim}
The intuition behind Claim~\ref{claim:u5} is that \reply{$u_{-3}$} changes if \reply{$u_0$} is smaller or greater than \reply{$u_1$} ($2$ choices), whereas \reply{$u_{-4}$} changes with the value of \reply{$u_0$} ($n-2$ choices). A detailed proof is given in Appendix~\ref{app:claim_u5}.


We refer to this encoding procedure as the \emph{horizontal VT encoder} and is denoted by $\VTabs{n}$.

\paragraph{Vertical VT encoder $\VTcds{n}$}Consider the message \mbox{$\mathbf{b} = (b_1,\dots,b_{k'}) \in \{1,\dots,n-1\}^{k'}$} to be encoded into the vector $\mathbf{v}$ of length $n' = n-\ell-1$. Here $ n'= n-\ell -1 = k'+5$. The vector $\mathbf{v} \in \Sigma_n^{n'}$ is encoded similarly to $\mathbf{u}$ except for the ordering of the data part and the parity part. 
\begin{enumerate}
    \item Let $v_1= b_1$ and $v_i = v_{i-1}+b_i \mod n$ for $i=1,\dots,k'$. 
    \item The symbol $v_{k'+1}$ can take an arbitrary value up to the constraints explained next. 
    \item We let $v_{k'+2}$ and $v_{k'+3}$ be the $n$-ary values of two length $\log n$ complement alternating sequences. 
    \item Computing the signature $\mathbf{s}'$ of the vector $(v_1,\dots,v_{k'+1})$, we let $v_{k'+4}$ be equal to $\sum_{i=1}^{k'+1}(i-1)s'_i \mod (k'+1)$.
    and $v_{k'+5} = \sum_{i=1}^{k'+1} v_i \mod n$.
    \item We choose $v_{k'+1}$ to be different from $v_{k'}$ and $v_{k'+2}$ such that $v_{k'+3}\neq v_{k'+4}$ and $v_{k'+4}\neq v_{k'+5}$. By Claim~\ref{claim:u5}, such a value of $v_{k'+1}$ always exists.
\end{enumerate}

We refer to this encoding procedure as the \emph{vertical VT encoder} and is denoted by $\VTcds{n}$.

\subsection{Encoder}
We are now ready to explain our explicit encoder for the \codename code construction. The encoder takes as input \mbox{$n_1 + n_2 +n_3$} bits and encodes them as follows, where 
\begin{align*}
    n_1 & = n^2 - 2n - 9\log n - (2n-\log n -11)\log \left(n\right) - 8, \\
    n_2 & = \left\lfloor(n-5) \log (n-1)\right \rfloor, \\
    n_3 & =\left\lfloor(n-\log n -6) \log (n-1)\right \rfloor. 
\end{align*}
\begin{enumerate}
    \item The first $\left\lfloor(n-5) \log (n-1)\right \rfloor$ bits are encoded using the horizontal VT encoder $\VTabs{n}$ and we let $\bfU$ be the binary array representation of the resulting vector.
    \item The next $\left\lfloor(n-\log n -6) \log (n-1)\right \rfloor$ bits are encoded using the vertical VT encoder $\VTcds{n}$ and let $\bfV$ be the binary array representation of the transpose of the resulting vector.
    \item The first bit of the alternating sequence representing \reply{$u_{-2}$} is repeated in the second column in the first and second row below $\bfU$. This bit is shown as $a$ in Figure~\ref{fig:encoding_exp}. The first bit of the alternating sequence representing $v_{n-\ell+1}$ is repeated in the second column in the third and fourth row below $\bfU$. This bit is shown as $b$ in Figure~\ref{fig:encoding_exp}.
    \item The alternating sequences representing \reply{$u_{-2}$} and \reply{$u_{-1}$} are extended by $1$ bit each. This bit is then repeated in the row below. Those bits are shown in Figure~\ref{fig:encoding_exp} as $c$ and $d$, respectively.
    \item The remaining $n^2 - 2n - {9\log n} - (2n-\log n -11)\log \left(n\right) - 8$ bits are systematically distributed in the $n\times n$ array outside of $\bfU$, $\bfV$, the positions of the parity check bits, and the eight reserved bits (shown in Figure~\ref{fig:encoding_exp}).
    \item The parity check bits are then computed as the respective column-wise and row-wise sums of all the bits.
\end{enumerate}

\subsection{Decoder}
The decoder works exactly the same as explained in Section~\ref{subsec:dec} where the alternating sequence is now the third column of $\bfU$ rather than the last column of $\bfU$ (even after either \reply{$u_{-2}$} or \reply{$u_{-1}$} 
is deleted). For completeness, we explain the subtle details of decoding deletions in $\bfU$ and $\bfV$. The insertion case follows similarly. 

The decoder first examines the received version of $\bfU$. To check whether a row of $\bfU$ is deleted, the decoder checks the alternating sequences (or one of them if the other is deleted). If a row is deleted, the alternating sequence must have a run of length $2$, unless the first row is deleted. If no run of length $2$ exists, the decoder simply counts the length of the alternating sequence to check if the first row is deleted. The decoder is guaranteed to count the exact length of the alternating sequence thanks to extending the sequence by one bit and repeating that bit. If a row of $\bfU$ is deleted, the decoder uses the row parity check to recover the value of the deleted row.

After checking for (and correcting) deleted rows in $\bfU$, the decoder checks for deleted columns. If both alternating sequences are not deleted and are in their correct positions, then the deleted column is in the systematic data part. The decoder uses the detailed decoding of \cite{tenengolts1984nonbinary} to recover the value and the position of the deleted column. If both alternating sequences are not deleted and are not in their correct positions, then the deleted column happened in the first two columns that are function of the systematic data part. The decoder can then recompute the parity part from the systematic data part and recover the index of the deleted column. In case one of the alternating sequences is deleted, then the decoder needs to know whether \reply{$u_{-2}$} or \reply{$u_{-1}$} 
is deleted. To that end, the decoder verifies the first bit of the non deleted alternating sequence with the bit in the second column and first row below $\bfU$. Thus, the decoder recovers the index of the deleted column. 

$\bfV$ is decoded similarly, except that the decoder would have recovered the index and value of the deleted column.

\subsection{Redundancy}
The redundancy of the explicit code is given by 
\begin{align*}
    R_{\text{explicit}} & = n^2-(n_1 + n_2 +n_3) \\
    & =  2n + 9\log n + (2n-\log n -11)\log \left(n\right) + 8 \\
    & ~~ - \left\lfloor(n-5) \log (n-1)\right \rfloor\\
    & ~~ -  \left\lfloor(n-\log n -6) \log (n-1)\right \rfloor\\
    & \stackrel{(a)}{<} 2n + 9\log n + (2n-\log n -11)\log \left(n\right) + 8 \\
    & ~~ - (n-5) \log (n-1) + 1\\
    & ~~ - (n-\log n -6) \log (n-1) + 1
\end{align*}
In inequality~(a) we used the inequality $\lfloor a \rfloor > a - 1$. 

Using the fact that $(2n-\log n -11) \log \left(\frac{n}{n-1}\right)$ is less than $2n \log \left(\frac{n}{n-1}\right)$ which is less than or equal to $2\log 2e = 2+ 2\log e$, we can write
\begin{align*}
    R_{\text{explicit}} 
    & = 2n + 9\log n + (2n-\log n -\reply{10})\log \left(\frac{n}{n-1}\right) + 10 \\
    & < 2n + 9\log n + 12 + 2\log e.
\end{align*}

%% file: Crisscross_explicit.tex
\begin{tikzpicture}


\definecolor{color0}{rgb}{0.12156862745098,0.466666666666667,0.705882352941177}
\definecolor{color1}{rgb}{1,0.498039215686275,0.0549019607843137}
\definecolor{color2}{rgb}{0.172549019607843,0.627450980392157,0.172549019607843}
\definecolor{color3}{rgb}{0.83921568627451,0.152941176470588,0.156862745098039}

\draw[pattern color=color0,pattern=vertical lines] (0,0) rectangle (6,-1.5);
\draw[pattern color=color1,pattern=horizontal lines] (4.5,-1.5) rectangle (6,-5.8);
\draw[pattern color=color2,pattern=north west lines] (0,-1.5) rectangle (0.2,-5.8);
\draw[pattern color=color2,pattern=north east lines] (0,-1.5) rectangle (0.2,-5.8);
\draw[pattern color=color3,pattern=north west lines] (0,-6) rectangle (6,-5.8);
\draw[pattern color=color3,pattern=north east lines] (0,-6) rectangle (6,-5.8);

\node[rectangle, fill = white, inner sep = 0pt] at (2.5, -0.7) {$\mathbf{U}\in \VTabs{n}$};
\node[rectangle, fill = white, inner sep = 0pt, rotate = 270] at (5.25, -3.5) {$\mathbf{V}\in \VTcds{n}$};;
\node[circle, fill = white, inner sep = 0pt] at (0.1,-3.5) {$\mathbf{p}_c$};
\node[circle, fill = white, inner sep = 0pt] at (2.5,-5.9) {$\mathbf{p}_r$};



 \node[inner sep = 0pt, anchor = north east, font = \scriptsize] at (0.6,-1.55) {$\begin{array}{c}a\\a\\b\\b\end{array}$};
\node[inner sep = 0pt, anchor = north east, font = \scriptsize] at (0.8,-1.55) {$\begin{array}{c}c\\c\end{array}$};
\node[inner sep = 0pt, anchor = north east, font = \scriptsize] at (1,-1.55) {$\begin{array}{c}d\\d\end{array}$};

\draw [decorate,decoration={brace,amplitude=10pt,mirror,raise=4pt},yshift=0pt, color = gray]
(-0.2,0) -- (-0.2,-1.5) node [black,midway,xshift=-1cm, color = gray] {\footnotesize
$\log n$};

\draw [decorate,decoration={brace,amplitude=10pt,mirror,raise=4pt},yshift=0pt, color = gray]
(-0.2,-1.5) -- (-0.2,-6) node [black,midway,xshift=-1.2cm, color = gray] {\footnotesize
$n-\log n$};

\draw [decorate,decoration={brace,amplitude=4pt,raise=1pt, mirror},yshift=0pt, color = gray]
(0,-5.8) -- (0,-6) node [black,midway,xshift=-0.3cm, color = gray] {\footnotesize
$1$};

\end{tikzpicture}

%% file: conclusion.tex

This paper considers the problem of criss-cross insertion/deletion in an $n\times n$ array. We have shown that every $(t)$-criss-cross deletion correcting code is a $(t)$-criss-cross insertion correcting code by extending the equivalence between insertion and deletion correcting codes from the one-dimensional case to the considered two-dimensional case. 

We derived a bound which shows that the redundancy of any \crisscross deletion/insertion correcting code is bounded from below by $2n-3+2 \log n$ for $n\geq 41$. We then constructed \codename code. This code can correct a single row and single column deletion in an $n\times n$ array. The redundancy of the \codename code is bounded from above by $2n + 4 \log n+7+2\log e$ bits. We have presented an explicit decoder for correcting deletions and insertions with this \codename code. We also modified this code construction to an explicit construction that has an explicit \emph{encoder} and an explicit decoder. The explicit encoder is based on systematic VT codes and comes at the expense of increasing the redundancy of the code by $5\log n +5$ bits.

\reply{In this work, we have considered deletions of one row {\em and} one column. Although our \codename code can correct a more general type of deletions, our bound on the redundancy and the equivalence proof do not directly hold in the more general model. Thus, as a future research direction, we are interested in investigating the case where any combination of $t_r$ rows and $t_c$ columns, such that $t_r+t_c$ is equal to a predetermined constant $t$, can be deleted or inserted. Another open problem of interest is also the case of mixed errors in which any $t_c$ columns may be deleted or inserted and any $t_r$ rows may be inserted or deleted.
We expect the techniques presented in this work to provide valuable insights on solving the more general problem. Preliminary results can be found in~\cite{welter2021multiple}.}

%% file: claim_ins.tex
 We prove that for any two arrays $\bfX_1, \bfX_{t+1}\in \sigmatnq$, $\iball_{t}(\bfX_1)\cap \iball_{t}(\bfX_{t+1})\neq \emptyset$ if and only if there exist $t-1$ arrays $\bfX_2,\dots,\bfX_{t}$ such that $\iball_{1}(\bfX_i)\cap \iball_{1}(\bfX_{i+1})\neq \emptyset$ for all $1\leq i \leq t$.
 
 We prove the ``if'' part by induction. The proof of the ``only if'' part follows similarly and is omitted.
 
 \paragraph{Base case} We need to show that if $\iball_{1}(\bfX_1)\cap \iball_{1}(\bfX_{2})\neq \emptyset$ then $\iball_{1}(\bfX_i)\cap \iball_{1}(\bfX_{i+1})\neq \emptyset$ for all $i=1$ which follows from the assumption.
 
\paragraph{Induction step}
Assume the property holds for $t\in [n-2]$ and we show that the property holds for $t+1$. Let $\bfX_1, \bfX_{t+2}$ be such that $\iball_{t+1}(\bfX_1)\cap \iball_{t+1}(\bfX_{t+2})\neq \emptyset$. Then, there exists $\bfXi{1}_1,\bfXi{1}_{t+1}$ resulting from a criss-cross insertion of $\bfX_1$ and $\bfX_{t+2}$, respectively, such that $\iball_{t}(\bfXi{1}_1)\cap \iball_{t}(\bfXi{1}_{t+1})\neq \emptyset$. Thus, according to the induction hypothesis, there exist $t-2$ arrays $\bfXi{1}_1,\dots,\bfXi{1}_{t}$ that satisfy $\iball_{1}(\bfXi{1}_i)\cap \iball_{1}(\bfXi{1}_{i+1})\neq \emptyset$ for all $1\leq i \leq t$.

According to Theorem~\ref{theorem:equiv}, there exist $t$ arrays $\bfX_2,\dots,\bfX_{t+1}$ such that for all $2\leq i \leq t+1$, $\bfX_i \in \dball(\bfXi{1}_{i-1})\cap \dball(\bfXi{1}_{i})$. Therefore, it holds that for $1\leq i \leq t+1$,
\begin{equation*}
    \bfXi{1}_i \in \iball_1(\bfX_{i})\cap \iball_1(\bfX_{i+1}).
\end{equation*}
This completes the ``if'' part of the proof. \hfill $\blacksquare$

%% file: cor_asymp.tex
We want to prove that when $n$ goes to infinity the redundancy of a criss-cross deletion correcting code is bounded from below by $2n-2+2\log_q n$.

To that end, we redefine a good array $\bfX$ to have a deletion ball greater than or equal to $n^2/2$. From Claim~\ref{claim:necessarily-good-arrays} we know that if an array $\bfX$ has more than $n/\sqrt{2}$ good rows and $n/\sqrt{2}$ good columns, then $\bfX$ is good.

Following the same steps of Lemma~\ref{lem:nbbad} we can bound the number of bad arrays (following this new definition) as
\begin{align*}
    \lvert \cB_n \rvert &\leq 2 \sum _{j=n-\frac{n}{\sqrt{2}}+1}^n \binom{n}{j} (b_n) ^j q^{n \cdot (n-j)} 
    \leq \sqrt{2}n 2^n  (2 n^2 )^n q^{\frac{n^2}{\sqrt{2}} - n}  \\
      & \leq \sqrt{2} q^{\frac{1}{\sqrt{2}}n^2 - n+\log_q(n) +n\log_q(4n^2)}\\
      &  \leq \sqrt{2} q^{\frac{1}{\sqrt{2}}n^2 - n+\log_2(n) +n\log_2(4n^2)}\\
    & \leq \sqrt{2} q^{n^2-3n},
\end{align*}
where $b_n \triangleq 3 \binom{n}{2}$ and the last inequality holds for $n \geq 54$.

Following the same steps of Theorem~\ref{thm:non-asymp-upper_bound}, we can bound the number of good arrays as $\lvert \cC_\cG\rvert \leq \frac{q^{n^2-1}}{\frac{n^2}{2}}$. We can now write
\begin{align*}
    \lvert \cC \rvert & = \lvert \cC_\cG \rvert +\lvert \cC_\cB \rvert \nonumber \\
    & \leq \lvert \cC_\cG \rvert +\lvert \cB_n \rvert \nonumber \\
    & \leq \frac{q^{(n-1)^2}}{\frac{n^2}{2}} + \sqrt{2} q^{n^2-3n} \nonumber \\
    & = \frac{q^{n^2}}{q^{2n-1}\cdot \frac{n^2}{2}}\left(1 +\frac{n^2}{\sqrt{2} \cdot q^{n+1}} \right)\\
    & \approx \frac{q^{n^2}}{q^{2n-1}\cdot \frac{n^2}{2}},
\end{align*}
where the last inequality is an asymptotic statement.

This concludes the proof. \hfill $\blacksquare$

%% file: cons_claims.tex
\begin{IEEEproof}[Proof of Claim~\ref{claim:rect}]
Remember that we have defined $\ell =\log n$. We first show that the redundancy of $\cU_\bot$ is given by
\begin{align*}
    R_{\cU_\bot} & = (n-\log n -1) \log \left( \frac{n}{n-1}\right).
\end{align*}
Recall that $\cU_\bot$ is defined as the set of all $n\times n$ arrays in which any two consecutive columns, from column $1$ to $n-\ell$, are different when restricted to the first $\ell$ entries, i.e.,

\begin{align*}
  \cU_\bot &\triangleq \left\{\bfX:
  \bfX_{[\ell],j} \neq \bfX_{[\ell],j+1}, \quad  j\in [n-\ell-1] \right\}.
\end{align*}
  
We count the number of arrays that satisfy those constraints. The first $\ell$ entries of the first column can take \reply{$2^\ell$} different values. For every other column from $2$ to $n-\ell$, the first $\ell$ entries can take \reply{$2^\ell-1$} different values because they have to be different from the entries of the column before. All other entries have no constraints and can take $2^{n^2 - \ell(n-\ell)}$ values. We can then write,
\begin{align*}
    |\cU_\bot| & = 2^\ell (2^\ell-1)^{n-\ell-1}2^{n^2 - \ell(n-\ell)}\\
    & = 2^{n^2} 2^{-(n-\ell -1)\ell}(2^\ell - 1)^{n-\ell -1}\\
    & = 2^{n^2} (1-2^{-\ell})^{n-\ell-1}.
\end{align*}
Thus, the redundancy can be computed as
\begin{align*}
    R_{\cU_\bot} & = n^2 - \log |\cU_\bot| \\
     & = -(n-\log n -1) \log \left(1 - \frac{1}{n}\right)\\
     & = (n-\log n -1) \log \left(\frac{n}{n-1}\right).
\end{align*}

To complete the proof we need to show that %
\begin{align*}
    R_{\cV_\bot} & = (n-\log n -2) \log \left( \frac{n}{n-1}\right) + 1.
\end{align*}

Recall that $\cV_\bot$ is defined as the set of all $n\times n$ arrays in which the entry $X_{\ell +1,n}$ is fixed to a predetermined value and any two consecutive rows, from row $\ell+1$ to $n-1$, are different when restricted to the last $\ell$ entries, i.e.,

\begin{align*}
  \cV_\bot & \triangleq \left\{\bfX:
  \begin{aligned}
  &\bfX_{i, [n-\ell+1,n]} \neq \bfX_{i+1, [n-\ell+1,n]} , \quad \ell<i<n-1 \\
  &{X_{\ell+1,n}} \equiv \ell \mod 2\\
  \end{aligned}\right\}.
\end{align*}

We count the number of arrays that satisfy those constraints. The last $\ell$ entries of row $\ell+1$ can take $2^{\ell-1}$ different values, because $X_{\ell +1,n}$ is predetermined. For every other row from $\ell+2$ to $n-1$, the last $\ell$ entries can take $2^\ell-1$ different values because they have to be different from the entries of the row before. All other bits have no constraints and can take $2^{n^2 - \ell(n-\ell-1)}$ values. We can then write,
\begin{align*}
    |\cU_\bot| & = 2^{\ell-1} (2^\ell-1)^{n-\ell-2}2^{n^2 - \ell(n-\ell-1)}\\
    & = 2^{n^2} 2^{-(n-\ell -2)\ell}(2^\ell - 1)^{n-\ell -2}2^{-1}\\
    & = 2^{n^2} (1-2^{-\ell})^{n-\ell-2}2^{-1}.
\end{align*}

The redundancy can then be computed as
\begin{align*}
    R_{\cU_\bot} & = n^2 - \log |\cU_\bot| \\
     & = -(n-\log n -2) \log \left(1 - \frac{1}{n}\right) +1 \\
     & = (n-\log n -2) \log \left(\frac{n}{n-1}\right) + 1.
\end{align*}

\end{IEEEproof}

Next we prove Claim~\ref{claim:square}, i.e. we show that the redundancy of $\cS_\cap$ is upper bounded by
\begin{equation*}
    R_{\cS_\cap} < \log n + 5.
\end{equation*}

\begin{IEEEproof}[Proof of Claim~\ref{claim:square}]
Recall that $\cS_\cap$ is defined as the set of $n\times n$ arrays in which the $\ell \times \ell$ sub array ending at the last bit of the first row of the original array has distinct consecutive columns, distinct consecutive rows, the last row fixed to a predetermined value and the first $4$ bits of the second to last column are also predetermined. $\cS_\cap$ also guarantees that the first column of the $\ell \times \ell$ sub array is different from the $\ell$ entries of column $n-\ell$ and similarly to the last row., i.e.,
\begin{equation*}
  \cS_\cap \triangleq \left\{\bfX:
  \begin{aligned}
  &\bfX_{[\ell],j} \neq \bfX_{[\ell],j+1} , \quad n-\ell\leq j<n,\\
  &\bfX_{i,[n-\ell+1,n]} \neq \bfX_{i+1,[n-\ell+1,n]}, i\in[\ell] ,\\
  &\bfX_{[4],n-1} = [0000]^T,\\
  &\bfX_{[\ell],n} = [010101\cdots]^T\\
  \end{aligned}\right\}.
\end{equation*}

Let $\cS_{c,r}$ be the set of arrays that have different consecutive columns and different consecutive rows. $\cS_\cap$ is the intersection between $\cS_{c,r}$ and the set of all arrays that have the first $\ell$ entries of the last column for an alternating sequence and the first $4$ entries of the second to last columns fixed to $0$. We shall prove in the sequel that $|\cS_{c,r}| > 2^{\ell^2-1}$. Once we have this bound, we can write
\begin{equation}
    |\cS_\cap| \geq \dfrac{|\cS_{c,r}|}{2^\ell 2^4} > \dfrac{2^{\ell^2}}{2^\ell 2^5}.\label{eq:cardinality}
\end{equation}

The first inequality follows from the fact that fixing the last column to a predetermined value reduces the number of arrays in $\cS_\cap$ by at most $2^\ell$ arrays and fixing $4$ bits of the second to last column reduces the number of arrays by at most $2^4$.

Therefore, using~\eqref{eq:cardinality} we have
\begin{align*}
    R_{\cS_\cap} & = \ell^2 - |\cS_\cap|
     \leq \ell + 5
     < \log n + 5.
\end{align*}

The remainder of the proof is to show that $|\cS_{c,r}| \geq 2^{\ell^2-1}$. We start by showing that the number of $\ell \times \ell$  arrays is lower bounded by $2^{\ell^2-1}$. This means that with one bit of redundancy we can guarantee the constraints on the rows and columns. 

To that end, we count the number of arrays that have at least two identical consecutive columns. Let $j$ and $j+1$, \reply{$j= n-\ell, \dots, n-1$}, be the indices of two identical consecutive columns. Column $j$ can take $2^\ell-1$ possible values and column $j+1$ can only take one value. Not imposing any constraints on the other $(\ell -2)$ columns, each column can have $2^\ell$ values and we have $(\ell -1)$ possible values for $j$. Therefore, the number of arrays having at least two identical consecutive columns is
$(\ell -1) (2^{\ell}-1)(2^\ell)^{\ell-2}.$

Following the same counting argument, the number of $\ell \times \ell$ arrays that have at least two identical consecutive rows is $(\ell -1) 2^{\ell}(2^\ell)^{\ell-2}$.

The number of arrays in $\cS_{c,r}$ is lower bounded by the total number of $\ell \times \ell$ arrays minus the number of arrays that have at least two identical consecutive columns and minus the number of arrays that have at least two identical consecutive rows. Thus, we can write
\begin{align}
    |\cS_{c,r}| & \geq 2^{\ell^2} - 2 (\ell - 1) (2^{\ell}-1)(2^\ell)^{\ell-2} \nonumber\\
    & > 2^{\ell^2} - 2 (\ell - 1) 2^{\ell}(2^\ell)^{\ell-2} \label{eq:ineq1}\\
    & \geq 2^{\ell^2 - 1}.\label{eq:ineq2}
\end{align}

The inequality in~\eqref{eq:ineq1} follows from
\begin{align}
    2 (\ell - 1) (2^{\ell}-1)(2^\ell)^{\ell-2} < 2 (\ell - 1) 2^{\ell}(2^\ell)^{\ell-2}, \label{eq:obvious}
\end{align}
which is true because $2^{\ell}-1<2^\ell$. The inequality in~\eqref{eq:ineq2} follows from
\begin{align} 
     2(\ell - 1) 2^{-\ell} \leq \dfrac{1}{2} \label{eq:easy}.
\end{align}
which is equivalent to \reply{$4(\ell -1) \leq 2^{\ell}$} and is true for all \reply{$\ell \geq 3$}. 
\end{IEEEproof}

%% file: u5_claim.tex
    We prove that for a vector \reply{$\mathbf{u} = (u_{-4},\dots,u_{n-5})\in \Sigma_n^n$}, there exists a value of \reply{$u_0$} such that the following holds: \begin{enumerate}
    \item \reply{$u_1,\dots,u_{n-5}$} can take arbitrary values such that any two consecutive symbols are different.
    \item \reply{$u_{-2}$} and \reply{$u_{-1}$} are the $n$-ary representation of two complement binary alternating sequences i.e., \reply{$u_{-2} = - u_{-1}$}, of length $\log n$ each.
    \item \reply{$u_{-4} = \sum_{i=0}^{n-5} u_i \mod n$}.
    \item \reply{$u_{-3}$} is equal to $$\displaystyle \sum_{i=1}^{n-4}(i-1)s_i \mod(n-4)$$ where, $\mathbf{s}=(s_1,\dots,s_{n-4})$ is the signature vector computed as $s_1 = 1$ and
    \begin{equation*}
      s_i = \begin{cases}
      1 & \hfill \text{ if } \reply{u_{i-1}\geq u_{i-2}}\\ 
      0 & \hfill \text{ otherwise.}
      \end{cases}
    \end{equation*}
    
    \item \reply{$u_0$} is chosen such that \reply{$u_{-4}\neq u_{-3}$}, \reply{$u_{-3}\neq u_{-2}$}, \reply{$u_{-1}\neq u_0$} and \reply{$u_0\neq u_1$}.
\end{enumerate}

Let \reply{$u_{-2},u_{-1}$} and \reply{$u_1,\dots,u_{n-5}$} be fixed. The symbol \reply{$u_{-3}$} can take two different values depending whether the chosen \reply{$u_0$} is less than or equal to \reply{$u_1$} or not. The value of \reply{$u_{-4}$} can take $n$ different values depending on the value of \reply{$u_0$}. Therefore, we start by ensuring that \reply{$u_{-3}$} is different than \reply{$u_{-2}$}, i.e., we choose if \reply{$u_0\leq u_1$} or \reply{$u_0>u_1$}. Assume that \reply{$u_0\leq u_1$}. Once \reply{$u_{-3}$} is fixed, we must choose a given value of \reply{$u_0$} such that \reply{$u_0\neq u_{-1}$} that makes \reply{$u_{-4}$} different than \reply{$u_{-3}$}. Notice that there is a one-to-one mapping between the value of \reply{$u_0$} and the value of \reply{$u_{-4}$}. Thus, since \reply{$u_{-1}$} is a large number, as long as \reply{$u_{-3} \geq 2$}, \reply{$u_0$} has at least two options ($0$ and $1$) out of which at least one satisfies all the aforementioned requirements. However, if \reply{$u_{-3} = 1$}, \reply{$u_0$} must be equal to $0$. In this case, if \reply{$u_{-4}$} is equal to \reply{$u_{-3}$} (then \reply{$u_0$} must be non zero) we switch the symbols \reply{$u_{-2}$} and \reply{$u_{-1}$} so that \reply{$u_0$} can now be greater than \reply{$u_1$} and has more than two different options that satisfy the aforementioned requirements. A similar argument holds for the case where \reply{$u_0> u_1$}. \hfill $\blacksquare$